\documentclass[%
 reprint,
 amsmath,amssymb,
 aps,
pra,
]{revtex4-2}

\usepackage{amsmath,amssymb}
\usepackage{bm}
\usepackage{graphicx}
\usepackage{dcolumn}
\usepackage{tabularx}
\usepackage{enumitem}
\usepackage{threeparttable}
\usepackage{booktabs}
\usepackage{cancel}
\usepackage[normalem]{ulem}
\usepackage{xcolor}
\usepackage{physics}

\graphicspath{{images/}}
\usepackage{orcidlink}
\usepackage{hyperref}
\hypersetup{
    colorlinks=true,
    citecolor=blue,
    linkcolor=blue,
    urlcolor=blue
}
\begin{document}

\title{\textbf{$r$-mode stabilization in rotating hyperon-rich neutron stars\\and its implications for GW190814}}

\author{Athira S.\orcidlink{0009-0005-3818-7186}}
\email{p23ph0002@iitj.ac.in}
\author{Monika Sinha\orcidlink{0000-0002-3080-9949}}
\thanks{Corresponding author}
\email{ms@iitj.ac.in}

\affiliation{$^{}$ Indian Institute of Technology Jodhpur, Jodhpur 342037, India}

\begin{abstract}
The GW190814 event, involving a black hole of mass $22.2$–$24.3 M_{\odot}$ and a compact object of mass $2.50$–$2.67 M_{\odot}$, challenges our understanding of the mass gap between the heaviest neutron stars and the lightest black holes. If the secondary is a neutron star exceeding $2.5 M_{\odot}$, hyperons are likely to appear in its core, softening the equation of state. Rapid rotation can offset some of this softening, enabling higher maximum masses, but it may simultaneously excite the Chandrasekhar–Friedman–Schutz $r$-mode instability. Bulk viscosity arising from nonleptonic weak interactions in hyperonic matter provides an efficient damping mechanism that can stabilize such configurations. In this work, we investigate the combined effects of rotation, thermal evolution, and hyperon-induced bulk viscosity on the stability of massive neutron stars. We demonstrate a direct connection between the suppression of $r$-mode instabilities and the long-term dynamical stability of hyperon-rich stars, offering a plausible interpretation of the GW190814 secondary as a rapidly rotating, hyperon-rich neutron star rather than a low-mass black hole. Our unified framework extends beyond previous studies restricted to static equations of state or extreme viscous damping assumptions, providing new insights into the stability of massive, exotic neutron star configurations.

\end{abstract}

\maketitle

\section{Introduction}
\label{intro}
The first multimessenger detection of a BNS merger, GW170817 \cite{LIGOScientific:2017vwq, Drout:2017ijr, Cowperthwaite:2017dyu}, constrained the maximum mass and radius distribution of neutron stars \cite{Rezzolla:2017aly, Annala:2017llu, Bauswein:2017vtn, Shibata:2019ctb, Margalit_2017, Radice_2018, Ruiz:2017due, Most:2018hfd, Raithel:2019uzi, Shibata_2019, Dietrich:2020efo, Nathanail:2021tay}. Much of this information has so far been extracted from the inspiral gravitational-wave (GW) signal; the postmerger phase, however, is expected to deliver even deeper insights into the equation of state (EOS) at supranuclear densities \cite{Stergioulas_2011, Bauswein:2011tp, Takami:2014zpa, Bernuzzi:2015rla, PhysRevD.93.124051, PhysRevLett.128.161102}, especially in scenarios involving a possible phase transition to deconfined quark matter \cite{Most:2018eaw, Most_2020, Weih:2019xvw, Tootle:2022pvd, Bauswein:2018bma, Blacker:2020nlq, Liebling:2020dhf, Prakash:2021wpz, Fujimoto:2022xhv, Ujevic:2022nkr}. In contrast to the GW170817 event, the GW190814 event, involving a 22.2–24.3 $M_\odot$ black hole and a 2.50–2.67$M_\odot$ compact object, highlights unresolved questions about the neutron star--black hole ``mass gap." Its secondary component is either the heaviest neutron star (NS) or the lightest black hole (BH) ever found in a binary compact object system, and the source has the most uneven mass ratio yet measured with gravitational waves \cite{LIGOScientific:2020zkf}. Previous works have explored both the upper \cite{Godzieba:2020tjn} and lower bounds on the maximum mass of the secondary component. As mentioned in the work of Most \textit{et al.}, the formation rates of such binary compact systems are uncertain. Still, all channels require dense stellar environments and are insensitive to whether the low-mass companion is a NS or BH. Assuming a mass gap of $\sim 5M_\odot$, the secondary must be a NS. If the secondary component is a NS with a mass exceeding \(2.5~M_\odot\), it is quite natural to expect the appearance of exotic matter in its core. However, the emergence of such exotic components generally softens the EOS, thereby reducing the maximum attainable mass. This issue can be mitigated by considering stellar rotation, since rapidly rotating neutron stars containing exotic matter in their cores can sustain higher masses. 

However, in the case of rotating neutron stars, the $r$-mode oscillation instability may arise, rendering the star unstable. Neutron stars exhibit various oscillatory modes driven by restoring forces such as pressure, gravity, and the Coriolis effect. Among these, inertial \emph{r} modes restored by the Coriolis force in rotating stars are particularly important because of their susceptibility to the Chandrasekhar-Friedman-Schutz (CFS) gravitational-wave-driven instability~\cite{Jyothilakshmi:2022hys, Ofengeim:2019fjy, Chatterjee:2006tk}. This instability occurs when gravitational radiation amplifies the mode energy more rapidly than viscous damping can dissipate it, potentially limiting the neutron star’s rotation rate~\cite{Chatterjee:2006tk} and accounting for the observed absence of pulsars spinning close to their Keplerian frequencies (e.g., the 716 Hz PSR~J1748$-$2446ad)~\cite{Dong:2025roh, Chatterjee:2006tk}.

Bulk viscosity, arising from pressure-density phase lags in perturbed matter, plays a crucial role in damping $r$-mode instabilities. Building on the pioneering estimates of bulk viscosity by Alford et al.~\cite{Alford:2017rxf}, subsequent studies have explored this mechanism in hot and dense nuclear matter under various conditions, including matter with trapped \cite{alford2021bulk, PhysRevD.100.103021} and untrapped neutrinos \cite{Alford:2023gxq, PhysRevC.109.015805, Alford:2021lpp, PhysRevC.100.035803}, with some incorporating muons in addition to the usual nucleons and electrons, and others extending the analysis to exotic phases such as hyperonic~\cite{2021PhRvC.103d5810A} and quark matter~\cite{Alford:2024tyj, CruzRojas:2024etx, Hern_ndez_2024}. In neutron star cores, exotic phases such as hyperonic matter are expected to appear at densities exceeding $\sim 2$–$3$ times the nuclear saturation density. At these higher densities, the Fermi momenta increase, allowing heavier hadrons to populate the system~\cite{Ofengeim:2019fjy, Chatterjee:2006tk, Glendenning:1991es}. Hyperons introduce nonleptonic weak interactions, which enhance bulk viscosity by several orders of magnitude compared to nucleon-only matter~\cite{Jyothilakshmi:2022hys, Ofengeim:2019fjy, Chatterjee:2006tk}. The prediction of millisecond-timescale damping has further motivated the inclusion of weak-interaction–driven bulk viscosity in neutron star merger simulations~\cite{PhysRevD.107.103032, Most_2024, Chabanov_2025, Espino:2023dei, Chabanov:2023abq, Radice_2022, Zappa_2023, Most_2021}. If the star is made of hyperons, the presence of hyperon bulk viscosity can effectively damp $r$-mode oscillations, rendering the star stable. Therefore, the companion in the GW190814 event may plausibly be a rotating hyperon-rich neutron star.

Our work aims to establish a novel connection between the thermal and rotational evolution of massive neutron stars with hyperonic compositions and their long-term dynamical stability, in contrast to earlier studies that primarily focused on static equations of state \cite{Dexheimer_2021, Biswas:2020xna, Fattoyev:2020cws} or invoked extreme viscous damping mechanisms to suppress $r$-mode instabilities in rapidly rotating configurations \cite{ANDERSSON_2001, Zhou_2021, papazoglou2017rmodeconstraintsneutronstar}.

The next Section, Sec.~\ref{sec:matter}, outlines the matter model employed in this work. In Sec.~\ref{sec:bv}, we briefly discuss the bulk viscosity arising in hyperonic matter, followed by an analysis of the attenuation of the $r$-mode due to bulk viscosity in Sec.~\ref{sec:rmode}. The results obtained using the equations of state considered in this study are presented in Sec.~\ref{sec:results}. Finally, the conclusions are summarized in Sec.~\ref{sec:summary}.

\section{Matter}\label{sec:matter}
We consider the matter is composed of baryon octet $(n,\, p,\, \Lambda,\, \Sigma^{+},\, \Sigma^{0},\, \Sigma^{-},\, \Xi^{0},\, \Xi^{-})$ together with leptons $(e^-, \mu^-)$. We employ a relativistic field theoretical model to describe charge-neutral and $\beta$-equilibrated hadronic matter.  The baryon-baryon interaction is mediated through the exchange of the isoscalar scalar meson $\sigma$, the isoscalar vector mesons $\omega, \phi$, and the isovector vector $\rho$ meson \cite{Schaffner:1993nn, Schaffner:1995th}. Accordingly, the Lagrangian density is written as
\begin{equation}
\begin{aligned}
\mathcal{L}_B 
&= \sum_B \bar{\psi}_B 
\Big[ 
    i\gamma_\mu \partial^\mu - m_B 
    + g_{\sigma B}\sigma 
    - g_{\omega B}\gamma_\mu \omega^\mu 
    - g_{\phi B}\gamma_\mu \phi^\mu \\
&\quad - g_{\rho B}\gamma_\mu \vec{\tau}_B \cdot \vec{\rho}^\mu \Big] \psi_B + \frac{1}{2}\left( 
    \partial_\mu \sigma \, \partial^\mu \sigma 
    - m_\sigma^2 \sigma^2 
\right) - U(\sigma) \\
&\quad - \frac{1}{4} \omega_{\mu\nu} \omega^{\mu\nu} 
+ \frac{1}{2} m_\omega^2 \omega_\mu \omega^\mu - \frac{1}{4} \vec{\rho}_{\mu\nu} \cdot \vec{\rho}^{\mu\nu} 
+ \frac{1}{2} m_\rho^2 \vec{\rho}_\mu \cdot \vec{\rho}^\mu \\
&\quad
- \frac{1}{4} \phi_{\mu\nu} \phi^{\mu\nu} 
+ \frac{1}{2} m_\phi^2 \phi_\mu \phi^\mu .
\end{aligned}
\end{equation}

The baryons are represented by the Dirac spinor $\psi_B$ with vacuum mass $m_B$ 
and isospin operator $\vec{\tau}_B$. The tensors $\omega_{\mu\nu}$, $\phi_{\mu\nu}$ and $\rho_{\mu\nu}$ denote the field strength tensors for the $\omega$, $\phi$ and $\rho$ mesons, respectively. The scalar self-interaction term \cite{Boguta:1977xi} is given by
\begin{equation}
U(\sigma) = \frac{1}{3} g_2 \sigma^3 + \frac{1}{4} g_3 \sigma^4 .
\end{equation}
This computation is performed within the mean-field approximation \cite{Serot:1984ey}. The quantities $\sigma$, $\omega_0$, $\rho_{03}$, and $\phi_0$ denote the mean values of the corresponding meson fields. Accordingly, we replace the meson fields by their expectation values, and the meson field equations take the form as given.

\begin{eqnarray}
m_\sigma^2 \sigma = \frac{-\partial U}{\partial \sigma} +  \sum_B g_{\sigma B}n_B^s \\ 
m_\omega^2 \omega_0 = \sum_B g_{\omega B}n_B \\
m_\phi^2 \phi_0 = \sum_B g_{\phi B}n_B \\
m_\rho^2 \rho_{03} = \sum_B g_{\rho B}I_{3B}n_B 
\end{eqnarray}
The baryon scalar and vector number densities are given by 
\begin{eqnarray}
n_{S}^B = \frac{2J_B + 1}{2\pi^2} \int_0^{k_{F_B}} \frac{m_B^{*}}{\sqrt{k^2 + m_B^{*2}}} k^2 dk \\
n_B = (2J_B + 1) \frac{k_{F_B}^3}{6\pi^2},
\end{eqnarray}
where $J_B$, $k_{F_B}$ and $I_{3B}$ are the spin, Fermi momentum and the third component of the isospin projection.\\
The effective mass of a baryon $B$ is given by
\begin{equation}
m_B^{*} = m_B - g_{\sigma B}\,\sigma,
\end{equation}
and the corresponding chemical potential is
\begin{equation}
\mu_B = \sqrt{k_{F_B}^2 + m_B^{*2}}
+ g_{\omega B}\,\omega_0 
+ g_{\phi B}\,\phi_0 
+ I_{3B}\, g_{\rho B}\,\rho_{03}.
\end{equation}
Since the hadronic phase is required to be charge neutral, the total charge density is
\begin{equation}
Q = \sum_B q_B n_B - n_e - n_\mu = 0 ,
\end{equation}
with $q_B$ being the electric charge of baryon $B$, and $n_e$, $n_\mu$ the number densities 
of electrons and muons, respectively.
The chemical equilibrium condition for each species is 
\begin{equation}
    \mu_i=b_i\mu_n - q_i\mu_e,
\end{equation}
where the chemical potentials of the $i$th baryon, neutrons and electrons are given as $\mu_i, \mu_n$, and $\mu_e$. The two conserved charges, baryon number and electric charge of the  $i$th baryon are given as $b_i$ and $q_i$.

The total energy density is given by  
\begin{equation}
\begin{aligned}
\varepsilon &= 
\tfrac{1}{2} m_{\sigma}^{2} \sigma^{2} 
+ \tfrac{1}{3} g_{2} \sigma^{3} 
+ \tfrac{1}{4} g_{3} \sigma^{4} 
+ \tfrac{1}{2} m_{\omega}^{2} \omega_{0}^{2} 
+ \tfrac{1}{2} m_{\rho}^{2} \rho_{03}^{2} \\
&\quad + \tfrac{1}{2} m_{\phi}^{2} \phi_{0}^{2} 
+ \sum_{B} \frac{2 J_{B} + 1}{2\pi^{2}} 
\int_{0}^{k_{F_{B}}} \sqrt{k^{2} + {m_{B}^{*}}^{2}} \, k^{2} \, dk \\
&\quad + \sum_{l} \frac{1}{\pi^{2}} 
\int_{0}^{k_{F_{l}}} \sqrt{k^{2} + m_{l}^{2}} \, k^{2} \, dk \, ,
\end{aligned}
\end{equation}
and the pressure is given by 

\begin{equation}
\begin{aligned}
P &= 
- \tfrac{1}{2} m_{\sigma}^{2} \sigma^{2} 
- \tfrac{1}{3} g_{2} \sigma^{3} 
- \tfrac{1}{4} g_{3} \sigma^{4} 
+ \tfrac{1}{2} m_{\omega}^{2} \omega_{0}^{2} 
+ \tfrac{1}{2} m_{\rho}^{2} \rho_{03}^{2} \\
&\quad + \tfrac{1}{2} m_{\phi}^{2} \phi_{0}^{2} 
+ \sum_{B} \frac{2 J_{B} + 1}{6 \pi^{2}}
\int_{0}^{k_{F_{B}}} \frac{k^{4}}{\sqrt{k^{2} + {m_{B}^{*}}^{2}}} \, dk\\
&\quad
+ \sum_{l} \frac{1}{3 \pi^{2}}
\int_{0}^{k_{F_{l}}} \frac{k^{4}}{\sqrt{k^{2} + m_{l}^{2}}} \, dk \, .
\end{aligned}
\end{equation}

\section{Bulk viscosity}\label{sec:bv}

In neutron-star matter, bulk viscosity is induced by $r$-mode oscillations, which periodically compress and expand the fluid, perturbing the pressure $P$ and baryon density $n$ from their equilibrium values. Consequently, the instantaneous pressure and number density deviate from the thermodynamic pressure and equilibrium number density described by the equation of state~\cite{Lindblom:2001hd,2021PhRvC.103d5810A}.  

This mismatch drives the system out of chemical equilibrium, since the particle fractions (neutrons, protons, hyperons, etc.) no longer satisfy the $\beta$ equilibrium conditions~\cite{Chatterjee:2006hy, Jha:2010an}. Weak interaction processes such as nonleptonic hyperon reactions and modified Urca processes act to relax the chemical potential imbalance, thereby damping the oscillations~\cite{Chatterjee:2006hy, Lindblom:2001hd}. These reactions, in turn, modify the neutron number density $n_{n}$ in response to density perturbations in the system.

When chemical equilibrium is restored, the forward and reverse reaction rates, $\Gamma_{f}$ and $\Gamma_{r}$, become equal. However, under $r$-mode perturbations, this balance is disturbed, and the difference between the rates,  
\begin{equation}
\delta \Gamma = \Gamma_{f} - \Gamma_{r} \neq 0 .
\end{equation}
emerges as the system strives to reestablish equilibrium.  

In neutron stars, bulk viscosity arises when density (or volume) oscillations drive the stellar fluid out of $\beta$ equilibrium. The resulting chemical potential imbalance induces weak interaction processes that act to restore equilibrium on a characteristic relaxation timescale $\tau$.
The relaxation time for the reaction is
\begin{equation}
\frac{1}{\tau} = 
\frac{\delta\Gamma}{\delta\mu}\frac{\delta\mu}{n_B\delta x_n},
\label{eq:Relax}
\end{equation}
where $\delta \mu$, $n_B$ and $\Gamma$ are the chemical imbalance, the total baryon density and total reaction rate, respectively.

The extent of damping caused by bulk viscosity is measured through the real part of the bulk viscosity coefficient $\zeta$. The bulk viscosity coefficient can be expressed as
\begin{equation}
\zeta = -\frac{n^2_B\tau}{1+(\omega\tau)^2}\frac{\partial P }{\partial n_n}\frac{d\bar{x}_n}{dn_B},
\label{eq:zeta}
\end{equation}
and $\bar{x}_n$ is the neutron fraction in the equilibrium state where $\bar{x}_n = n_n/n_B$.\\
Here, the $r$-mode frequency $\omega$, observed in the corotating frame, is linked to the star's rotation frequency $\Omega$ to first order and is expressed as \cite{Andersson:2002ch, Papaloizou:1978zz,1981A&A....94..126P}
\begin{equation*}
    \omega=\frac{2m\Omega}{l(l+1)}.
\end{equation*}
Our analysis is concerned with the $l = m = 2$ $r$ mode.

\subsection{Nonleptonic reactions}
The relaxation times of weak interaction processes fall within a few orders of magnitude of the $r$-mode oscillation period and therefore serve as the dominant contributors to bulk viscosity. Among these, the most important processes are the non-leptonic weak interactions involving the lightest hyperon, $\Lambda$, since it appears at lower densities and reaches larger populations compared to heavier particles.
We have the maximum bulk viscosity contributing nonleptonic hyperon reaction given as
\begin{align}
n + p &\rightleftharpoons p + \Lambda  \label{eq:proc1} 
\end{align}
It sets the lower bound for the reaction rate and, in turn, establishes an upper bound on the bulk-viscosity coefficient. Although several processes contribute to the overall reaction rate \cite{Jha:2010an, vanDalen:2003uy, Nayyar:2005th}, we focus exclusively on this particular process, as it dominates the bulk-viscosity contribution and thus serves as the most relevant channel for our analysis \cite{2001PhRvL..86.1384J, 2001PhRvD..64h4003J,1969Ap&SS...5..213L,Haensel:2001em}.

For the reaction. \eqref{eq:proc1} the reaction rate is given as
\begin{equation}
    \delta\Gamma = \frac{1}{192\pi^3}\left\langle \left| \mathcal{M} \right|^2\right\rangle k_{F_{\Lambda}}(kT)^{2}\delta\mu,
\end{equation}
where $k_{F_\Lambda}$ is the Fermi momentum of $\Lambda$, k is the Boltzmann constant, T is the temperature and $\left\langle \left| \mathcal{M} \right|^2\right\rangle$ is the angle-averaged matrix element (squared and summed over spin states), given as \cite{Lindblom:2001hd}

\begin{widetext}
\normalsize
\begin{align}
\left\langle \left| \mathcal{M} \right|^2 \right\rangle
&= \frac{G_{F}^{2} \sin^{2}(2\theta_{C})}{15} \Bigg\{ 
120 (1-g_{np}^{2})(1-g_{p\Lambda}^{2}) m_{n} m_{p}^{2} m_{\Lambda} \nonumber  - 20 (1-g_{np}^{2})(1+g_{p\Lambda}^{2}) m_{n} m_{p} \big( 3\epsilon_{p}\epsilon_{\Lambda} - k_{F_{\Lambda}}^{2} \big) \nonumber \\[6pt]
&\quad - 10 (1+g_{np}^{2})(1-g_{p\Lambda}^{2}) m_{p} m_{\Lambda} 
\big( 6\epsilon_{n}\epsilon_{p} - 3k_{F_{n}}^{2} + k_{F_{\Lambda}}^{2} \big) \nonumber+ 2\big[(1+g_{np}^{2})(1+g_{p\Lambda}^{2}) + 4g_{np}g_{p\Lambda}\big] \nonumber \\[6pt]
&\qquad \times \Big[ 
5\epsilon_{p}\epsilon_{\Lambda} (6\epsilon_{n}\epsilon_{p} + 3k_{F_{n}}^{2} - k_{F_{\Lambda}}^{2} )+ k_{F_{\Lambda}}^{2} \big(10\epsilon_{n}\epsilon_{p} + 5k_{F_{n}}^{2} + 10k_{F_{p}}^{2} - k_{F_{\Lambda}}^{2}\big) 
\Big] \nonumber + \big[(1+g_{np}^{2})(1+g_{p\Lambda}^{2}) \\&- 4g_{np}g_{p\Lambda}\big] \times \Big[ 
10\epsilon_{n}\epsilon_{\Lambda} \big(6m_{p}^{2} + 3k_{F_{n}}^{2} + k_{F_{\Lambda}}^{2}\big) + k_{F_{\Lambda}}^{2}(
-20\epsilon_{p}^{2} + 15k_{F_{p}}^{2} - 3k_{F_{\Lambda}}^{2} + \frac{5(k_{F_{n}}^{2}-k_{F_{p}}^{2})^{2}}{(k_{F_{p}}^{2}-k_{F_{\Lambda}}^{2})}
\Big] 
\Bigg\},
\label{eq:MLambda}
\end{align}
\end{widetext}

where $G_{F}^{2}$ is the Fermi coupling constant, $\theta_{C}$ is the Cabibbo weak mixing angle, the quantities $g_{np}$ and $g_{p\Lambda}$ are the axial-vector couplings, $k_{F_{i}}$ denotes the Fermi momenta, and $\epsilon_i$ the particle energies \cite{Lindblom:2001hd}. The values we use, $G_{F} = 1.166 \times 10^{-11},\mathrm{MeV}^{-2}$ and $\sin \theta_{C} = 0.222$, are taken from the Particle Data Group \cite{ParticleDataGroup:2024cfk} and that of $g_{np}=-1.27$ and $g_{p\Lambda}=-0.72$ are obtained from $\beta$ decay of baryons at rest \cite{ParticleDataGroup:2024cfk}.

For reaction \eqref{eq:proc1}, the chemical potential imbalance is 
\begin{equation}
    \delta \mu \equiv \delta \mu_n - \delta \mu_\Lambda \, =  \sum_i (\alpha_{ni} - \alpha_{\Lambda i}) \delta n_i,
    \label{eq:delmu}
\end{equation}
where $i= n $ and $\Lambda$ for the matter we consider and $\alpha_{ij} = (\frac{\partial \mu_i}{\partial n_j})_{n_{k},k \neq j}$.
The number density of baryon species normalized to the total number density of baryons is $ x_i =  n_i/ n_B$, then we have $\delta x_i = \delta n_i/ n_B$, since $\delta n_B \equiv 0$. Thus, we obtain the expression
\begin{equation}
\frac{\delta \mu}{n_{B} \, \delta x_{n}} = \alpha_{nn} - \alpha_{\Lambda n} - \alpha_{n \Lambda} + \alpha_{\Lambda \Lambda}.
\label{eq:delmun}
\end{equation}
Here $\delta x_n = x_n - \Tilde {{x}_n}$ represents the deviation of the neutron fraction from its equilibrium value $\Tilde{x_n}$. 
The general expression for $\alpha_{ij}$ is given as 
\begin{align}
\alpha_{ij} 
&= \frac{\pi^2}{k_{F_i}\mu_i^*}\,\delta_{ij} 
- \frac{m_i^*}{\mu_i^*}
  \frac{\left(\tfrac{g_{\sigma i}}{m_\sigma}\right)
        \left(\tfrac{g_{\sigma j}}{m_\sigma}\right)
        \tfrac{m_j^*}{\mu_j^*}}{D} \nonumber \\[4pt]
&\quad + \frac{g_{\omega i} g_{\omega j}}{m_\omega^2} 
+ \frac{g_{\rho i} I_{3i}\, g_{\rho j} I_{3j}}{m_\rho^2}.
\label{eq:alphaij}
\end{align}

Here, D is expressed as 
\begin{equation}
    D = 1 + \sum_B \left(\frac{g_B}{m_\sigma}\right)^2 \frac{\partial n_S^B}{\partial m_B^*} + \frac{1}{m_\sigma^2} \frac{\partial^2 U}{\partial \sigma^2}.
\end{equation}
Equation \eqref{eq:delmun} is substituted in Eq. \eqref{eq:Relax} for the determination of the microscopic relaxation time ($\tau$). Using this, we find the bulk viscosity coefficient for the nonleptonic process from Eq.~\eqref{eq:zeta}.

\subsection{Urca reactions}
Along with the nonleptonic reaction, the leptonic Urca process also contributes to the relaxation time.
In dense neutron star matter, the direct Urca process is the most efficient neutrino emission channel, and it is given as \cite{Haensel:2000vz, 1992ApJ...390L..77P}
\begin{equation}
n \rightarrow p + e^- + \bar{\nu}_e , \qquad p + e^- \rightarrow n + \nu_e .
\end{equation}
The direct Urca process occurs only  when the following momentum condition is satisfied,
\begin{equation}
k_{F_n} < k_{F_p} + k_{F_e}.
\end{equation}
This condition implies that the proton and electron Fermi momenta and hence their number densities must be sufficiently large compared to that of neutrons. Therefore, the direct Urca channel operates exclusively in the cores of very massive neutron stars, where the central density is high enough to permit this threshold.  

For the direct Urca reaction, the reaction rate is given as \cite{Sinha:2008wb}
\begin{equation}
    \delta\Gamma = \frac{17}{240\pi}m_n^* m_p^*  \mu_l ( \left| \mathcal{M} \right|_d^2)_{\theta_{int}} (kT)^{4}\delta\mu.
\end{equation}
The subscript $\theta_{int}$ refers to the angle-integrated quantity. In degenerate matter, where particles occupy states close to their Fermi surfaces, the momenta and energies appearing in the integration are approximated by their values at the corresponding Fermi surfaces.

We have the squared matrix element of the reaction as \cite{Sinha:2008wb}
\begin{align}
\bigl|\mathcal{M}\bigr|_{d,\theta_{\mathrm{int}}}^{2}
&= G_{F}^{2} \cos^{2}\theta_{C}
\Big[
(1 + g_{np})^{2}
\left(1 - \frac{k_{F_n}}{m_{n}^{*}}\right)
\\[4pt]
&\quad + (1 - g_{np})^{2}
\left(1 - \frac{k_{F_p}}{m_{p}^{*}}\right)
\nonumber - (1 - g_{np}^{2})
\Big].
\label{eq:M_theta_int}
\end{align}

For the relaxation time equation \eqref{eq:Relax}, we have the following expression as 
\begin{equation}
\frac{\delta \mu}{n_{B} \, \delta x_{n}} = \alpha_{nn} - \alpha_{np} + \alpha_{pp} - \alpha_{pn} + \alpha_{ll}  , 
\end{equation}
where the change in the lepton chemical potential in response to a perturbation in the lepton number density $\alpha_{ll}$ is given as $\alpha_{ll}=\pi^2/k_{F_{l}} \mu_l$, where $l=e, \mu$. We evaluate the timescale ($\tau_U$) and bulk viscosity coefficient of the direct Urca process ($\zeta_U$) in the same manner as done for the nonleptonic case.

 The direct Urca process is kinematically allowed only when the proton fraction exceeds a critical threshold. If this condition is not satisfied, momentum conservation forbids the direct Urca reactions, and they are instead replaced by the modified Urca (mUrca) process, where an additional nucleon participates to conserve momentum \cite{Haensel:2001mw}. For neutron stars of lower mass, the proton fraction remains too small to satisfy this condition, rendering the direct Urca process forbidden. In addition to the direct Urca processes, the mUrca reactions play a significant role in neutron star matter in regions where the direct Urca is kinematically forbidden. In the outer cores of massive neutron stars, as well as in the cores of relatively low-mass stars ($M \lesssim 1.3$–$1.4 M_\odot$) \cite{1995A&A...297..717Y}, the direct Urca process is kinematically forbidden because momentum conservation cannot be simultaneously satisfied. The direct Urca process becomes kinematically forbidden when the particle momenta are such that,  
\begin{equation}
    k_{F_p} + k_{F_e} < k_{F_n} .
\end{equation}
Under these conditions, the dominant neutrino emission channel is the modified Urca process. Unlike the direct Urca process, which involves only a neutron and a proton in the weak interaction, the modified Urca requires the participation of an additional bystander nucleon to conserve momentum. The neutron branch gives the corresponding reactions 
\begin{equation}
    n + n \rightarrow n + p + e^- + \bar{\nu}_e , \qquad n + p + e^- \rightarrow n + n + \nu_e ,
\end{equation}
and the proton branch
\begin{equation}
n + p \rightarrow p + p + e^- + \bar{\nu}_e , \qquad p + p + e^- \rightarrow n + p + \nu_e .
\end{equation}
These processes are less efficient compared to the direct Urca channel, but they become the dominant neutrino emission mechanism in regimes where direct Urca is suppressed \cite{1995A&A...297..717Y, Haensel:2001mw}. 

 The following expression gives the bulk viscosity coefficient for the modified Urca process \cite{Lindblom:1998wf, PhysRevD.39.3804}
\begin{equation}
    \zeta_U = 6 \times 10^{-59} \epsilon^2 \omega^{-2} T^6.
    \label{eq:tau_u}
\end{equation}
where $\epsilon$, $\omega$, and $T$ are the energy density, angular velocity of the corotating frame and temperature, respectively.
Hence, Urca reactions are essential for describing the bulk viscosity in neutron stars, albeit at a reduced level compared to the nonleptonic hyperon processes. 

\section{$r$-mode Attenuation}\label{sec:rmode}
$r$ modes (or Rossby waves) are a distinct category of inertial modes that, according to Newtonian theory, are predominantly axial to leading order~\cite{Andersson:2002ch}. These modes arise from axial oscillations, with the Coriolis force acting as the restoring force~\cite{Kraav_2024}. Since the Coriolis force itself originates from stellar rotation, $r$ modes exist only in rotating stars. Consequently, their properties are intrinsically tied to the rotation rate, the mode frequency scales linearly with the angular velocity, and the oscillation pattern corresponds to large-scale circulation of fluid elements around the rotation axis~\cite{ANDERSSON_2001}. In the absence of rotation, the Coriolis force vanishes, and the $r$ modes cease to exist. Thus, $r$ modes represent global oscillations whose restoring mechanism and dynamical behavior are entirely governed by rotation. The associated oscillatory variations in pressure and density drive the system out of $\beta$ equilibrium, and the weak-interaction mediated relaxation of particle fractions converts oscillation energy into heat. This process produces dissipative bulk viscosity, which acts to suppress the growth of $r$ modes. In this work, we consider the mode with $l = m = 2$, which is the most significant $r$ mode capable of rendering rapidly rotating stars unstable and efficiently emitting gravitational waves through the CFS instability~\cite{Li:2023gtg}. The imaginary part of the $r$-mode damping time scale ($\tau_i$), represents the macroscopic timescale over which the $r$ mode is damped, and is given by~\cite{Lindblom:2001hd, Jha:2010an, 2008arXiv0806.3359S}
\begin{equation}
    \frac{1}{\tau_i} = -\frac{1}{2E}\frac{dE}{dt} \, ,
\end{equation}
where E is the perturbation energy given by 
\begin{equation}
    E = \frac{1}{2} \alpha^2 R^{-2} \int_0^R \epsilon(r)r^6dr
    \label{eq:E}
\end{equation}
with $\alpha$ the dimensionless $ r$-mode amp. Ilitude coefficient, R and $\epsilon$ the radius and the energy density profile of the star and dissipation rate 
\begin{equation}
    \frac{dE}{dt} = -4\pi \int_0^R \zeta(r)\, \left\langle \left| \vec{\nabla} . \delta\vec{v} \right|^2 \right\rangle \, r^2 dr.
    \label{eq:dE/dt} 
    \end{equation}
Here, the angle average of the square of the hydrodynamic expansion~\cite{Lindblom:1999yk} 
\begin{equation}
\left\langle \left| \vec{\nabla} . \delta\vec{v} \right|^2 \right\rangle \ = \frac{(\alpha \Omega)^2}{690}\left(\frac{r}{R}\right)^6 \left[1+0.86 (\frac{r}{R})^2\right] \left(\frac{\Omega^2}{\pi G \bar{\epsilon}}\right)^2
\label{eq:delv}
\end{equation}
with $G$, $\bar{\epsilon}$ are the gravitational constant and the mean energy density of a nonrotating star. For the present analysis, we find that the rotating models considered exceed the Tolman-Oppenheimer-Volkoff (TOV) maximum mass ($M_{TOV}$) \cite{1939PhRv...55..364T,1939PhRv...55..374O,1997A&A...328..274B}, so the mean energy density $\bar{\epsilon}$ is taken from the maximum-mass nonrotating model. This quantity serves only as a normalization factor in the hydrodynamic expansion. Its specific value has a negligible effect on the computed dissipation timescales, which are primarily determined by the radial dependence of the bulk-viscosity coefficient $\zeta(r)$ entering the radial integrals. The subscript $i=B$ for the nonleptonic process and $i=U$ for the Urca process.

Now, using Eqs. \eqref{eq:E}, \eqref{eq:dE/dt}, and \eqref{eq:delv}, together with the stellar density profile $\rho(r)$ and the bulk viscosity coefficient $\zeta(r)$ for the different damping mechanisms such as bulk viscosity due to nonleptonic processes (hyperon bulk viscosity) and bulk viscosity due to Urca reactions, we can determine the damping time scale for each process.

Gravitational radiation is known to destabilize $r$ modes in all rotating stars, providing a driving mechanism that amplifies these oscillations. On the other hand, viscous dissipation acts as a damping mechanism, counteracting this instability. Understanding the competition between these two effects requires particular attention to the hyperon bulk viscosity timescale, $\tau_{B(h)}$. The imaginary part of the $r$-mode frequency can then be expressed as
\begin{equation}
\frac{1}{\tau_r} = -\frac{1}{\tau_{GR}} + \frac{1}{\tau_{B}} + \frac{1}{\tau_{U}},
\end{equation}
where ~\cite{Lindblom:1998wf}
\begin{align}
\frac{1}{\tau_{GR}}
&= -\,\frac{32\pi G\,\Omega^{2l+2}}{c^{2l+3}}
\frac{(l-1)^{2l}}{\left[(2l+1)!!\right]^2}
\nonumber \\[4pt]
&\quad \times 
\left( \frac{l+2}{l+1} \right)^{2l+2}
\int_{0}^{R} \rho\, r^{2l+2}\, dr,
\label{eq:tau_gr}
\end{align}

where $\tau_{GR}$, $\tau_{B}$, and $\tau_{U}$ denote the timescales for gravitational radiation, hyperon bulk viscosity, and modified Urca bulk viscosity, respectively \cite{Lindblom:2001hd, Chatterjee:2006hy}.
The $r$ mode becomes stable when the imaginary part of its frequency vanishes, i.e.
\begin{equation}
\frac{1}{\tau_r(\Omega_c,T)} = 0,
\label{eq:1bytau}
\end{equation}
which defines the critical angular velocity of the star, $\Omega_c$.
The critical angular velocity $\Omega_c$ plays a fundamental role in the study of $r$-mode instability. It represents the threshold angular velocity at which the energy gained through gravitational radiation is exactly balanced by the energy lost due to viscous dissipation. This dissipation is primarily governed by bulk viscosity, which strongly depends on the internal temperature and composition of the star \cite{Chatterjee:2006hy, Jha:2010an}. Neutron stars rotating with angular velocities greater than $\Omega_c$ become unstable to $r$-mode oscillations, leading to the amplification of these modes and a subsequent spin down due to angular momentum loss via gravitational wave emission. Conversely, stars rotating more slowly than $\Omega_c$ remain stable, as viscous damping suppresses the growth of the $r$ modes \cite{Jyothilakshmi:2022hys,Jha:2010an}.
 
\section{Results}\label{sec:results}
\begin{figure}   
    \centering
    \includegraphics[angle=-90, width=9.5cm]{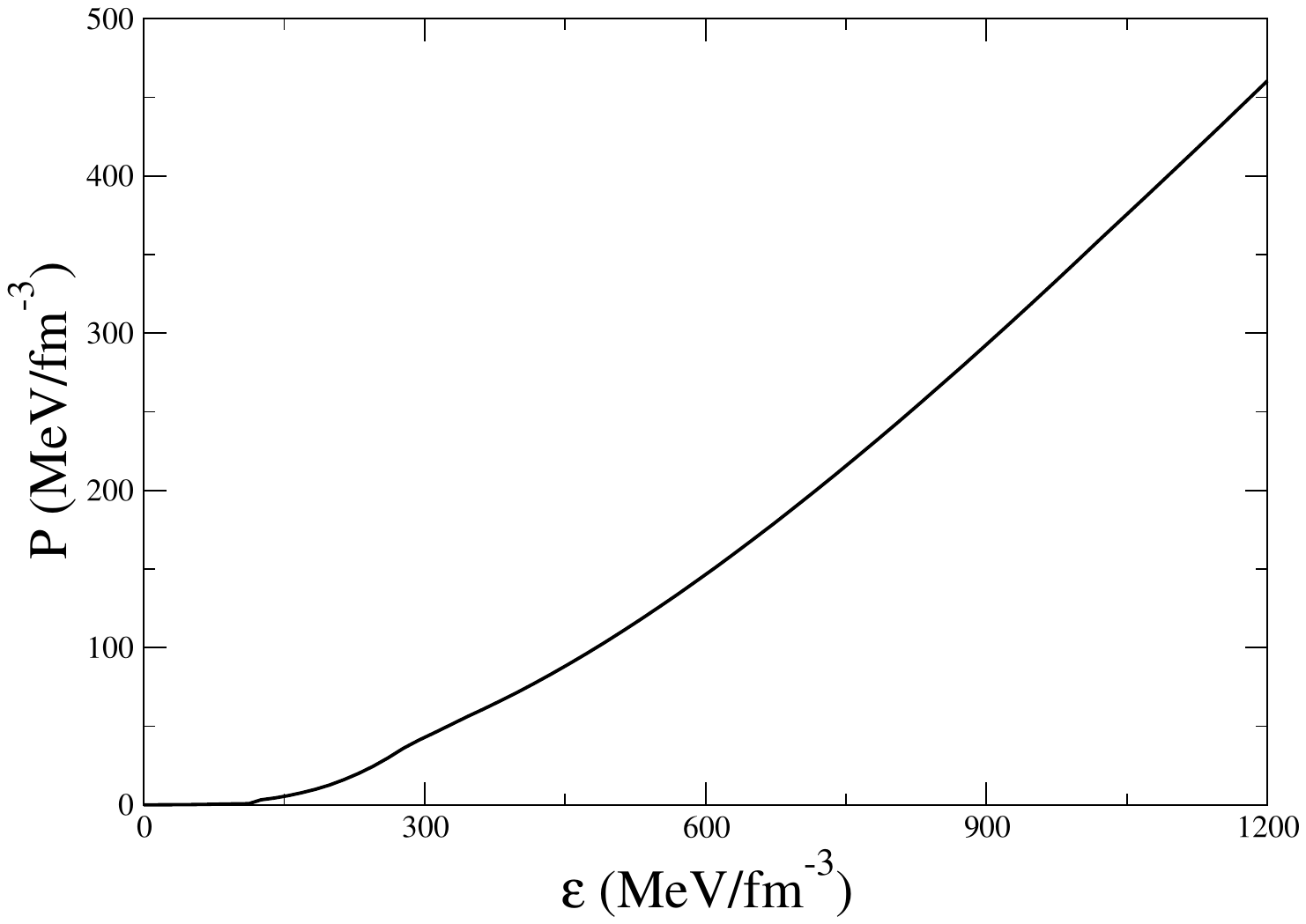}
    \caption{ Variation of pressure with energy density for the NL-SH parametrization.}
    \label{fig:PvsE}
\end{figure}

In the present work, we construct a NS with hyperons in the interior. For the matter calculation, we employ the NL-SH parametrization within the mean field approach to obtain the matter EOS. The EOS corresponding to this parametrization is shown in Fig.~\ref{fig:PvsE}.

In this parametrization, the masses of the particles are given in Table \ref{table:mass}. The nucleon couplings are given in Table \ref{table:nucleon_couplings}. For the hyperon-vector meson coupling constants, we adopt the SU(6) symmetry relations derived from the quark model, which are given as follows
\begin{align}
    \frac{1}{2} g_\omega \Lambda &= \frac{1}{2} g_{\omega \Sigma} = g_{\omega \Xi }= \frac{1}{3} g_{\omega N} ,\\
    2 g_\phi \Lambda &= 2 g_\phi \Sigma = g_\phi \Xi = -\frac{2\sqrt{2}}{3} g_{\omega N} ,\\
   g_{\rho \Lambda} &= 0 ,\\
    \frac{1}{2} g_{\rho \Sigma} &= g_{\rho \Xi }= g_{\rho N}. 
\end{align}
The scalar meson-hyperon couplings are determined self-consistently from the empirical potential depths of the hyperons $\Lambda$, $\Sigma$, and $\Xi$, considered as $-28$, $+30$, and $-14$ MeV, respectively, in nuclear matter. This approach ensures that the hyperon interactions are constrained both by underlying symmetry considerations and by available experimental and phenomenological inputs, thereby providing a more reliable description of dense matter.

\begin{table*}[t]
\centering
\caption{Mass values (in units of MeV) for the different particles in the NL-SH model.}
\resizebox{\textwidth}{!}{%
\begin{tabular}{ccccccccccc}
\hline
\hline
Model & $m_\sigma$ & $m_\omega$ & $m_\rho$ & $m_\phi$ & $m_N$ & $m_\Lambda$ & $m_\Sigma$ & $m_\Xi$ & $m_e$ & $m_\mu$ \\
\hline
NL-SH & 526.059 & 783.00 & 763.00 & 1020.00 & 939.00 & 1116.00 & 1193.00 & 1313.00 & 0.50 & 105.65 \\
\hline
\hline
\end{tabular}%
}
\label{table:mass}
\end{table*}

\begin{table}[t]
\centering
\caption{Numerical values of the nucleon couplings in the NL-SH model.}
\resizebox{\columnwidth}{!}{%
\begin{tabular}{cccccc}
\hline
\hline
Model & $g_{\sigma N}$ & $g_{\omega N}$ & $g_{\rho N}$ & $b_1$ & $c_1$  \\
\hline
NL-SH & 10.4444 & 12.9450 & 8.7660 & 6.9099 & -15.8337  \\
\hline
\hline
\end{tabular}%
}
\label{table:nucleon_couplings}
\end{table}

If we construct the static NS ($\Omega = 0$) without any exotic matter in the core of the star, then the maximum attainable mass is found to be $2.79~M_\odot$. However, for such a massive star, it is natural that exotic matter should appear inside the interior. If we construct a static NS with hyperons in its core, then the maximum attainable mass is found to be $2.35~M_\odot$, which is less than the estimated secondary star mass of the observed GW190814 event. However, if we consider the rotating hyperon star with rotational frequency $\Omega=5150$ s$^{-1}$, we achieve the maximum attainable mass  $M=2.5~M_\odot$, which is just above the lower limit of the estimated secondary star mass of the observed GW190814 event. Hence, the hyperon star with angular velocity more than $\Omega\approx5150$ s$^{-1}$ can be the secondary component of the binary merger event GW190814. 

Our hyperon-rich equations of state yield relatively large TOV maximum masses, exceeding several GW170817-based upper limits of \(M_{\rm TOV}\approx 2.17\,M_{\odot}\) \cite{Margalit_2017,Rezzolla:2017aly,Ruiz:2017due}, and even the more permissive bound of \(\sim 2.3\,M_{\odot}\) inferred under specific postmerger assumptions \cite{Shibata_2019}. We stress that these limits are model dependent, relying on assumptions about the remnant lifetime, rotational support, and spin evolution. In our models, the higher \(M_{\rm TOV}\) values originate from high-density stiffening driven by repulsive hyperonic interactions consistent with causality. Recent multimessenger studies further suggest that the upper bound on \(M_{\rm TOV}\) may be relaxed if longer-lived supramassive remnants are allowed \cite{Ai:2023ykc,Margalit:2022rde}.\\

However, rotating neutron stars are subject to $r$-mode instability. In the following, we demonstrate that this instability can be efficiently damped by bulk viscosity arising from hyperonic processes in dense matter. 

\begin{figure}[b]   
    \centering
    \includegraphics[angle=-90, width=9.5cm]{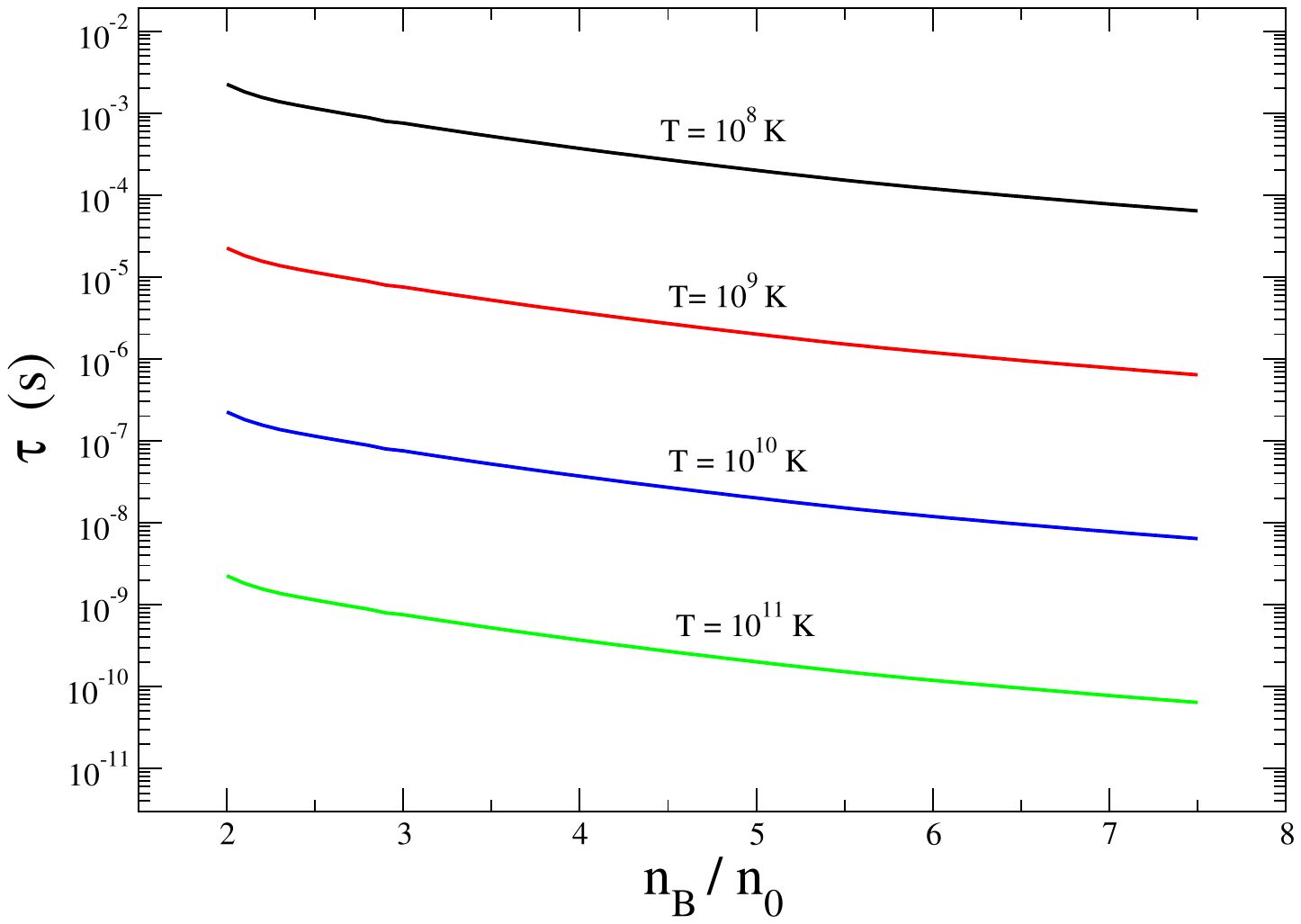}
    \caption{Relaxation time is plotted with normalized baryon density for the nonleptonic process for different temperatures.}
    \label{fig:TvsN(all)}
\end{figure}

Using the adopted parametrization, we compute the relaxation time corresponding to the hyperonic reaction in Eq.~\eqref{eq:proc1}. The variation of the relaxation time with normalized baryon density for different temperatures is shown in Fig.~\ref{fig:TvsN(all)}. The variation with temperature indicates that the relaxation time is strongly influenced by thermal effects: as the temperature increases, the relaxation time decreases. This trend is clearly illustrated in Fig.~\ref{fig:TvsN(all)}.

\begin{figure}[b]   
    \centering
    \includegraphics[angle=-90, width=9.5cm]{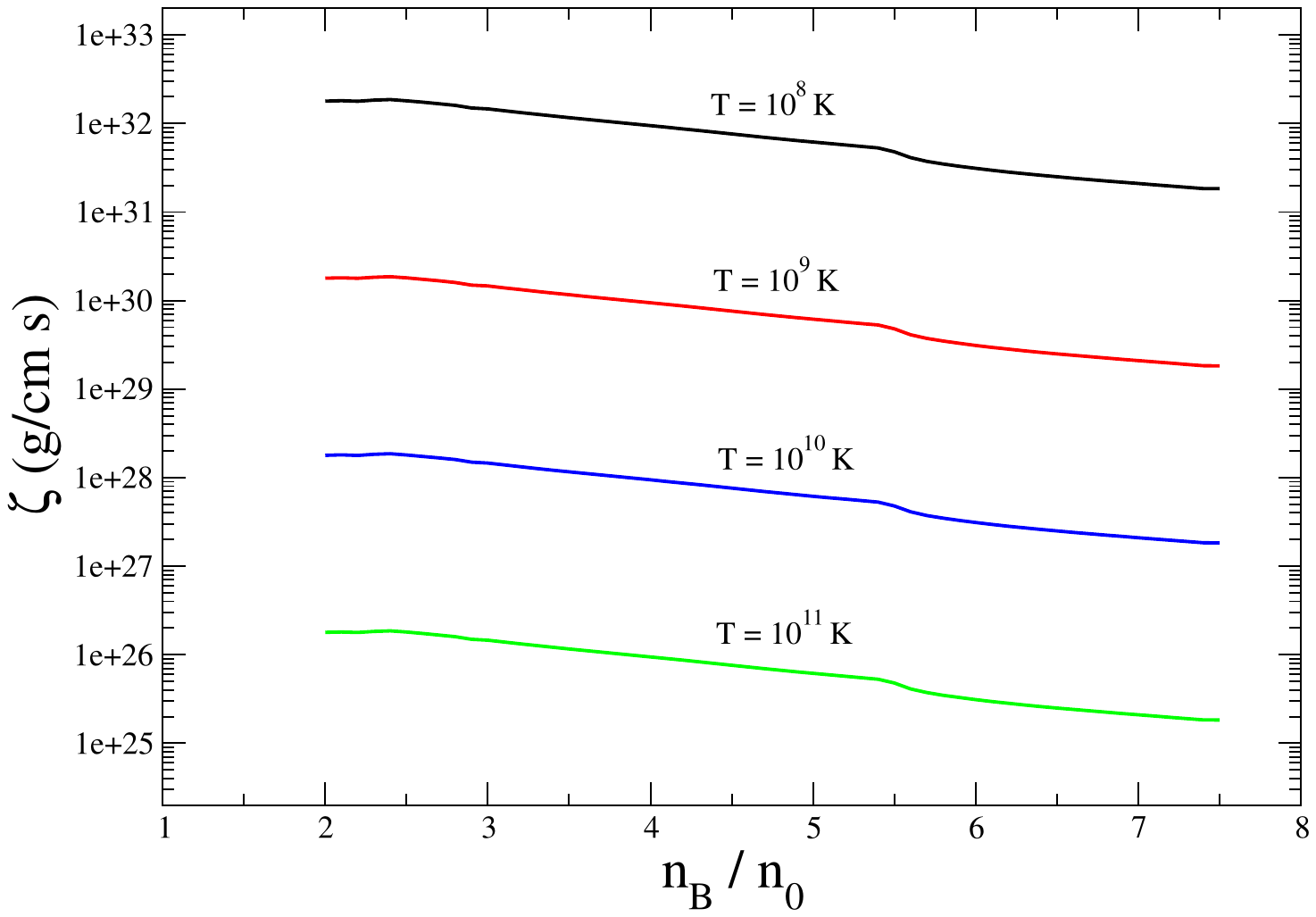}
    \caption{Bulk viscosity coefficient is exhibited as a function of normalized baryon density for different temperatures.}
    \label{fig:ZvsN(all)}
\end{figure}

From the relaxation time, we obtain the bulk viscosity coefficient. The variation of $\zeta$ with the normalized baryon density is shown in Fig.~\ref{fig:ZvsN(all)}. The temperature dependence reveals that the hyperonic bulk viscosity is largest at the lowest temperature. Specifically, as illustrated in Fig.~\ref{fig:ZvsN(all)}, the maximum bulk viscosity occurs at a temperature of the order of $10^{8}$ K among the cases considered. In the temperature regime considered in this study, the parameter $\omega \tau \ll 1$ is for the nonleptonic weak processes involving hyperons. This condition implies that the relaxation timescale of the system is much shorter than the oscillation period of the stellar fluid. This means that any deviation from chemical equilibrium induced by density oscillations is restored almost instantaneously through rapid nonleptonic reactions. As a result, the dissipation reaches a saturated regime, where the bulk viscosity becomes nearly independent of temperature. Therefore, the term containing $(\omega \tau)^2$ in the denominator of Eq.~\eqref{eq:zeta} can be neglected when calculating the hyperonic bulk viscosity.

\begin{figure}   
    \centering
    \includegraphics[angle=-90, width=9.5cm]{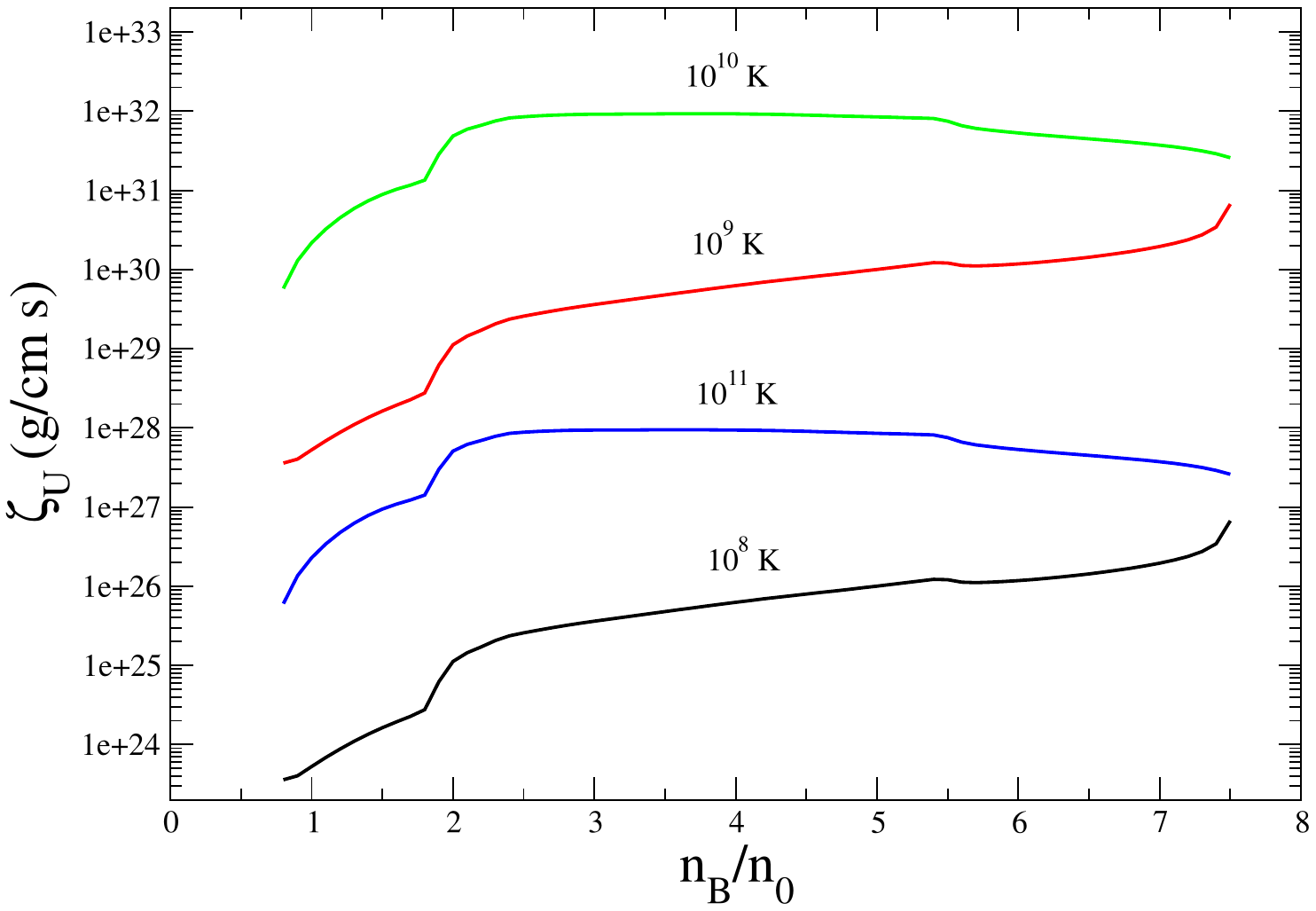}
  \caption{Bulk viscosity coefficient due to the Urca process is exhibited as a function of normalized baryon density for different temperatures.}
    \label{fig:ZurcavsN(all)}
\end{figure}

Figure~\ref{fig:ZurcavsN(all)} illustrates the variation of $\zeta$ arising from the Urca processes with the normalized baryon density for different temperatures. It is evident from the figure that the bulk viscosity increases steadily as the temperature rises from \(10^{8}\) to \(10^{10}\) K, reaching a maximum around \(10^{10}\) K. Beyond this point, at \(10^{11}\)K, the bulk viscosity exhibits a noticeable decline~\cite{Sinha:2008wb, Haensel:2000vz}. This characteristic temperature dependence results from the interplay between the timescale of the Urca reactions and the $r$-mode oscillation period of the stellar matter. For temperatures up to \(10^{10}\) K, the condition \(\omega\tau > 1\) holds, leading to a \(T^{4}\) dependence of the bulk viscosity. However, at higher temperatures around \(10^{11}\) K, where \(\omega\tau < 1\), the dependence reverses to \(T^{-4}\). At lower temperatures, the reaction rates are too slow to efficiently restore \(\beta\) equilibrium, resulting in weak viscous dissipation. As the temperature increases, the reaction rates accelerate, enhancing the energy dissipation and thereby increasing the bulk viscosity. The viscosity attains its peak when the reaction timescale becomes comparable to the oscillation period of the stellar mode. At higher temperatures (\(T \sim 10^{11}\) K), the reactions occur so rapidly that the matter remains nearly in instantaneous equilibrium throughout the oscillation cycle, thereby reducing the overall dissipation and causing the bulk viscosity to decrease again.
The bulk viscosity arising from the direct Urca processes is activated abruptly at the threshold density where the electron direct Urca process becomes possible (\( n_B = 0.219~\mathrm{fm^{-3}} \)). The appearance of muons reduces this threshold density, primarily because their presence increases the proton number density and Fermi momentum. However, muons simultaneously decrease the bulk viscosity associated with the electron direct Urca process by reducing the electron number density (\( n_e \)). At higher baryon densities, the total bulk viscosity experiences a second sharp increase (at \( n_B = 0.2774~\mathrm{fm^{-3}} \)) corresponding to the onset of the muon direct Urca process, in which muons themselves participate. The contribution of this muon Urca process to the total bulk viscosity surpasses that of the electron Urca process. Such behavior is anticipated because muons enable an extra nonequilibrium Urca process, increasing the viscosity of stellar matter \cite{Haensel:2000vz}.

\begin{figure}   
    \centering
    \includegraphics[angle=-90, width=9.5cm]{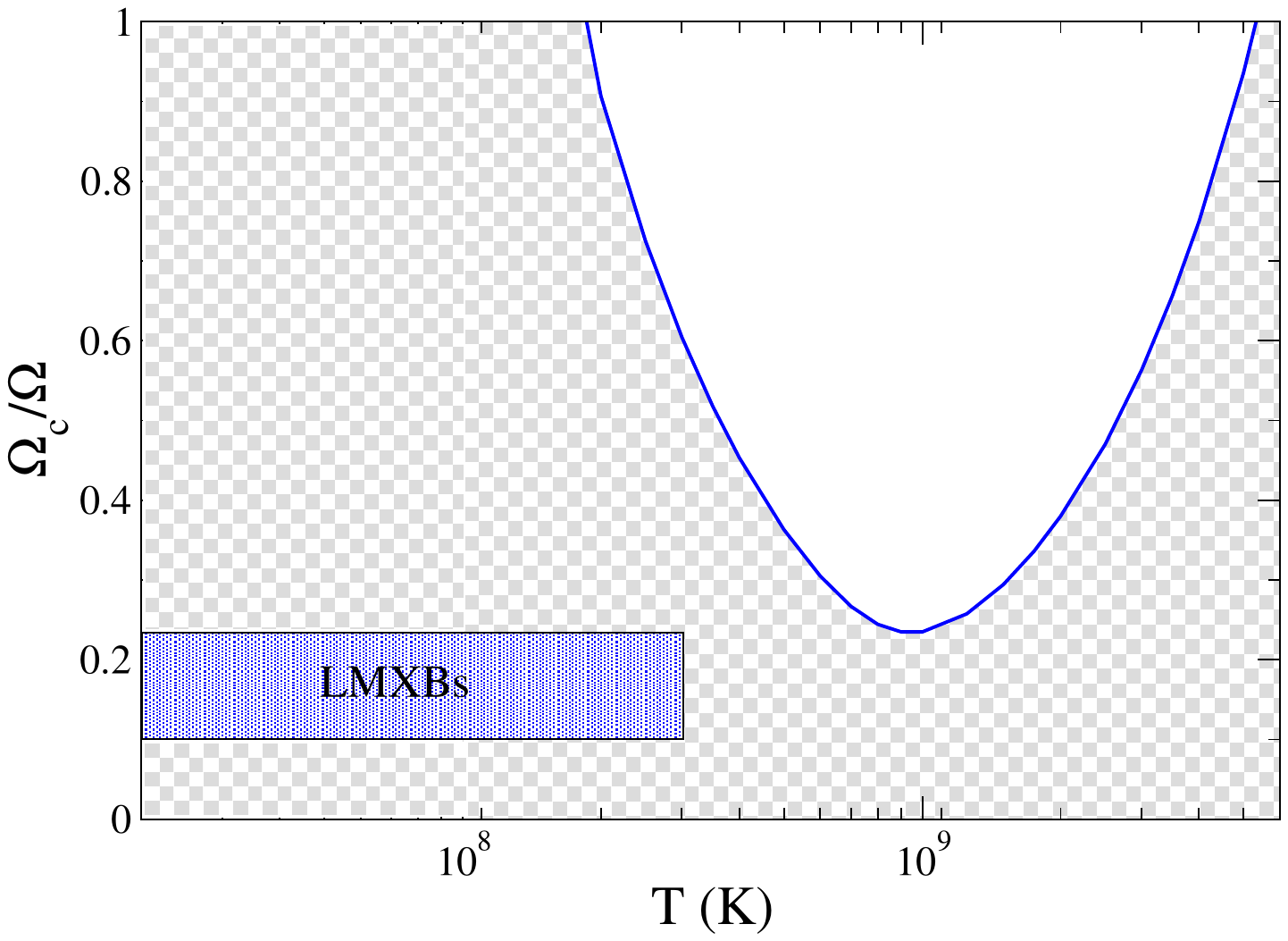}
    \caption{Critical angular velocities for the neutron star of mass $2.67 M_\odot$ with rotational frequency of $\Omega = 6510\,\mathrm{s^{-1}}$ are plotted as a function of temperature. The shaded region below the curve corresponds to the stable region, and the shaded box depicts the observed domain of LMXBs.}
    \label{fig:OvsT}
\end{figure}

Then we find the critical angular velocity $\Omega_c$ of the star below which the $r$-mode oscillation is damped. As $\zeta$ varies with temperature, $\Omega_c$ also varies with temperature. Hence, in the $\Omega_c-T$ plot, we find an instability window for rotating neutron stars. With the EOS as described above, the star with mass $2.67~M_\odot$, the upper limit of the estimated mass range for the GW190814, can be achieved with rotational frequency $\Omega = 6510 $ s$^{-1}$. We show the instability window for rotating neutron stars of mass $2.67~M_\odot$ in Fig.~\ref{fig:OvsT}. This is obtained by evaluating $\Omega_c$ as a function of temperature through solving Eq.~\eqref{eq:1bytau}. The instability window spans the temperature range from $1.86 \times 10^8$ K to $5.27 \times 10^{9}$ K. So the star with mass $2.67~M_\odot$ is stable against the gravitational wave radiation due to $r$-mode instability if its temperature is not in this range. If temperature is higher than $1.86 \times 10^8$ K and less than $5.27 \times 10^{9}$ K then depending on the rotational speed of the star the gravitational radiation drives the mode efficiently, with viscous damping unable to counteract it, indicating that a neutron star containing a hyperonic core loses its rotational frequency due to gravitational radiation loss in this regime. 

A neutron star with a gravitational mass in the estimated range of mass of the secondary component of GW190814, \(2.5-2.67\,M_{\odot}\) with high rotational speed can originate from the core-collapse supernova explosion of a massive progenitor star with an initial mass exceeding about \(8\,M_{\odot}\). At birth, such a newly formed protoneutron star is characterized by extremely high internal temperatures, typically in the range of \(10^{10}\)--\(10^{11}\,\mathrm{K}\) \cite{Glendenning:1997wn}, resulting from the rapid compression of matter during collapse. This temperature, however, decreases rapidly within the first few tens of seconds due to intense neutrino emission, which efficiently carries away thermal energy from the stellar interior. In this early stage, the star lies in the high-temperature regime of the $r$-mode instability curve, where enhanced viscous damping, primarily due to bulk viscosity, suppresses the growth of unstable oscillation modes, rendering the star stable against $r$-mode instabilities. As the neutron star cools further, continued neutrino emissions drive its temperature downward into the so-called $r$-mode instability window, where viscous damping becomes less effective compared to gravitational radiation-driven growth. In this regime, $r$-mode instabilities may develop, leading to angular momentum loss through gravitational wave emission and resulting in a temporary spin-down phase. The associated dissipation processes can also generate significant internal heating, which may transiently alter the cooling trajectory of the star. This unstable phase, however, is not permanent. As the temperature continues to decline, the star eventually exits the instability window and reenters a stable region of the $r$-mode plane, provided it cools through the critical temperature range sufficiently rapidly and retains adequate angular momentum. Once the instability window is traversed, the star attains a quasistable configuration. For our case a neutron star rotating with an angular velocity range of \(\Omega = 5150 - 6510 \,\mathrm{s^{-1}}\), for a model of mass range \(2.5-2.67 \,M_{\odot}\), corresponds to the estimated region in the parameter space where the star remains dynamically stable against $r$-mode oscillations. This suggests that the secondary component of GW190814 could indeed be a hyperonic neutron star, consistent with both theoretical predictions and the observed mass constraints. With an estimated rotational frequency range of approximately \(819.5-1039~\mathrm{Hz}\), corresponding to an angular velocity of \(5150-6510~\mathrm{s^{-1}}\), it would represent the most massive and fastest neutron star ever observed, thereby providing a remarkable link between dense matter physics and gravitational-wave observations.

In Fig.~\ref{fig:OvsT}, the shaded region below the curve represents the stable part and the shaded box depicts the observed domain of low-mass x-ray binaries (LMXBs), spanning temperatures between $2 \times 10^{7}$ and $3 \times 10^{8}$~K, and rotational frequencies from $300$ to $700$~Hz \cite{Jyothilakshmi:2022hys, 2004ApJ...605..830C, Huang:2009ue, Brown:2002rf}. The instability window shows perfect agreement with the data obtained from astrophysical observations.

\section{Summary and Conclusion}\label{sec:summary}

We address the problem of identifying the secondary component in the binary neutron star merger event GW190814. In this event, the secondary component has a mass in the range $2.50$ to $2.67~M_\odot$ \cite{LIGOScientific:2020zkf}. If the secondary object is a NS, then it is natural to expect that such a massive NS must contain exotic matter in its core. However, the appearance of exotic matter softens the equation of state, thereby reducing the maximum supported mass of a neutron star. In this scenario, only a rapidly rotating NS containing exotic matter would be able to sustain such a high mass. Rotating neutron stars, however, are susceptible to $r$-mode instabilities. These instabilities can be damped by bulk viscosity arising from the chemical imbalance created during oscillations in dense matter. Consequently, we propose that, for massive NSs, the presence of hyperons can generate sufficient bulk viscosity to suppress the $r$-mode instability.

For newly born neutron stars, rapid rotation is often accompanied by the emission of gravitational waves driven by $r$-mode oscillations. The $r$-mode instability acts as an efficient mechanism for angular momentum loss, resulting in a gradual spin down of the star and an accompanying increase in the central density. This rise in density modifies the internal composition of the star, making hyperon formation particularly sensitive to the rotation rate~\cite{Chatterjee:2006hy}. Therefore, the effect of hyperon bulk viscosity on the stability of rotating NSs becomes crucial.

The present analysis demonstrates that a NS with a gravitational mass in the range of \(2.5-2.67\,M_{\odot}\) can naturally evolve through the $r$-mode instability window during its cooling phase. As the star cools from its initially high birth temperature, it passes through the instability region and subsequently regains stability.

While our cold EOS supports uniformly rotating neutron stars up to $\approx 2.5{-}2.67\,M_{\odot}$, this exceeds GW170817-derived limits of $\lesssim 2.3\,M_{\odot}$ from kilonova/GRB modeling \cite{Margalit_2017,Rezzolla:2017aly,Ruiz:2017due,Shibata_2019}. These limits assume specific postmerger dynamics (differential rotation decay, fallback mass $\sim0.01{-}0.04\,M_{\odot}$, outflows) and carry $\sim0.1\,M_{\odot}$ systematic uncertainties across numerical relativity simulations. Thus, our equilibrium configurations remain in tension with, but not robustly excluded by, these model-dependent inferences.

By explicitly constructing the instability curve depicting the critical angular velocity as a function of temperature for hyperonic matter, and tracking the star’s evolutionary trajectory across this plane, we provide a self-consistent framework that links the microphysical composition of dense matter to macroscopic stability behavior. This approach offers a unified explanation for the existence and sustained stability of the secondary component in GW190814, even in the presence of exotic degrees of freedom, thereby extending beyond previous treatments that considered the thermal evolution, rotational dynamics, and composition of dense matter as largely independent phenomena.

\vspace{5pt}

\section{Acknowledgements}
The authors thank the anonymous referee for the valuable suggestions provided, which have greatly contributed to enhancing the quality of this manuscript. The authors acknowledge the financial support from the Science and Engineering Research Board (SERB), Department of Science and Technology, Government of India, through Project No. CRG/2022/000069. M.S. acknowledges the partial financial support from the DRDO through Project No. DGTM/ERIP/GIA/24-25/010/005.

\section{Data Availability}
No data were created or analyzed in this study.

\bibliographystyle{spphys}
\bibliography{references}

@article{Jyothilakshmi:2022hys,
    author = "Jyothilakshmi, O. P. and Krishnan, P. E. Sravan and Thakur, Prashant and Sreekanth, V. and Jha, T. K.",
    title = "{Hyperon bulk viscosity and r-modes of neutron stars}",
    eprint = "2208.14436",
    archivePrefix = "arXiv",
    primaryClass = "astro-ph.HE",
    doi = "10.1093/mnras/stac2360",
    journal = "Mon. Not. Roy. Astron. Soc.",
    volume = "516",
    pages = "3381--3388",
    year = "2022"
}

@article{Ofengeim:2019fjy,
    author = "Ofengeim, D. D. and Gusakov, M. E. and Haensel, P. and Fortin, M.",
    title = "{Bulk viscosity in neutron stars with hyperon cores}",
    eprint = "1911.08407",
    archivePrefix = "arXiv",
    primaryClass = "astro-ph.HE",
    doi = "10.1103/PhysRevD.100.103017",
    journal = "Phys. Rev. D",
    volume = "100",
    pages = "103017",
    year = "2019"
}

@article{Dong:2025roh,
    author = "Dong, Wenhao and Melatos, Andrew",
    title = "{Gravitational waves from r-mode oscillations of stochastically accreting neutron stars}",
    eprint = "2501.04968",
    archivePrefix = "arXiv",
    primaryClass = "astro-ph.HE",
    doi = "10.1093/mnras/staf033",
    journal = "Mon. Not. Roy. Astron. Soc.",
    volume = "537",
    pages = "650--660",
    year = "2025"
}

@article{Chatterjee:2006tk,
    author = "Chatterjee, Debarati and Bandyopadhyay, Debades",
    title = "{Exotic bulk viscosity and its influence on neutron star r-modes}",
    eprint = "astro-ph/0607005",
    archivePrefix = "arXiv",
    doi = "10.1007/s10509-007-9356-4",
    journal = "Astrophys. Space Sci.",
    volume = "308",
    pages = "451--455",
    year = "2007"
}

@article{LIGOScientific:2020zkf,
    author = "Abbott, R. and others",
    collaboration = "LIGO Scientific, Virgo",
    title = "{GW190814: Gravitational Waves from the Coalescence of a 23 Solar Mass Black Hole with a 2.6 Solar Mass Compact Object}",
    eprint = "2006.12611",
    archivePrefix = "arXiv",
    primaryClass = "astro-ph.HE",
    reportNumber = "LIGO-P190814",
    doi = "10.3847/2041-8213/ab960f",
    journal = "Astrophys. J. Lett.",
    volume = "896",
    pages = "L44",
    year = "2020"
}

@article{Papaloizou:1978zz,
    author = "Papaloizou, J. and Pringle, J. E.",
    title = "{Non-radial oscillations of rotating stars and their relevance to the short-period oscillations of cataclysmic variables}",
    doi = "10.1093/mnras/182.3.423",
    journal = "Mon. Not. Roy. Astron. Soc.",
    volume = "182",
    pages = "423--442",
    year = "1978"
}

@ARTICLE{1981A&A....94..126P,
       author = {{Provost}, J. and {Berthomieu}, G. and {Rocca}, A.},
        title = "{Low Frequency Oscillations of a Slowly Rotating Star - Quasi Toroidal Modes}",
      journal = {\aap},
         year = 1981,
        month = jan,
       volume = {94},
        pages = {126},
       adsurl = {https://ui.adsabs.harvard.edu/abs/1981A&A....94..126P}
}

@article{Andersson:2002ch,
    author = "Andersson, Nils",
    title = "{Gravitational waves from instabilities in relativistic stars}",
    eprint = "astro-ph/0211057",
    archivePrefix = "arXiv",
    doi = "10.1088/0264-9381/20/7/201",
    journal = "Class. Quant. Grav.",
    volume = "20",
    pages = "R105",
    year = "2003"
}

@article{Chatterjee:2006hy,
    author = "Chatterjee, Debarati and Bandyopadhyay, Debades",
    title = "{Effect of hyperon-hyperon interaction on bulk viscosity and r-mode instability in neutron stars}",
    eprint = "astro-ph/0602538",
    archivePrefix = "arXiv",
    doi = "10.1103/PhysRevD.74.023003",
    journal = "Phys. Rev. D",
    volume = "74",
    pages = "023003",
    year = "2006"
}

@article{Jha:2010an,
    author = "Jha, T. K. and Mishra, H. and Sreekanth, V.",
    title = "{Bulk viscosity in hyperonic star and r-mode instability}",
    eprint = "1002.4253",
    archivePrefix = "arXiv",
    primaryClass = "hep-ph",
    doi = "10.1103/PhysRevC.82.025803",
    journal = "Phys. Rev. C",
    volume = "82",
    pages = "025803",
    year = "2010"
}

@article{Boguta:1977xi,
    author = "Boguta, J. and Bodmer, A. R.",
    title = "{Relativistic Calculation of Nuclear Matter and the Nuclear Surface}",
    doi = "10.1016/0375-9474(77)90626-1",
    journal = "Nucl. Phys. A",
    volume = "292",
    pages = "413--428",
    year = "1977"
}

@article{Schaffner:1993nn,
    author = "Schaffner, Jurgen and Dover, Carl B. and Gal, Avraham and Greiner, Carsten and Stoecker, Horst",
    title = "{Strange hadronic matter}",
    reportNumber = "BNL-48889",
    doi = "10.1103/PhysRevLett.71.1328",
    journal = "Phys. Rev. Lett.",
    volume = "71",
    pages = "1328--1331",
    year = "1993"
}

@article{Schaffner:1995th,
    author = "Schaffner, Jurgen and Mishustin, Igor N.",
    title = "{Hyperon rich matter in neutron stars}",
    eprint = "nucl-th/9506011",
    archivePrefix = "arXiv",
    doi = "10.1103/PhysRevC.53.1416",
    journal = "Phys. Rev. C",
    volume = "53",
    pages = "1416--1429",
    year = "1996"
}

@article{Serot:1984ey,
    author = "Serot, Brian D. and Walecka, John Dirk",
    title = "{The Relativistic Nuclear Many Body Problem}",
    reportNumber = "ITP-740-STANFORD",
    journal = "Adv. Nucl. Phys.",
    volume = "16",
    pages = "1--327",
    year = "1986"
}

@article{Lindblom:2001hd,
    author = "Lindblom, Lee and Owen, Benjamin J.",
    title = "{Effect of hyperon bulk viscosity on neutron star r modes}",
    eprint = "astro-ph/0110558",
    archivePrefix = "arXiv",
    doi = "10.1103/PhysRevD.65.063006",
    journal = "Phys. Rev. D",
    volume = "65",
    pages = "063006",
    year = "2002"
}

@ARTICLE{2021PhRvC.103d5810A,
       author = {{Alford}, Mark G. and {Haber}, Alexander},
        title = "{Strangeness-changing rates and hyperonic bulk viscosity in neutron star mergers}",
      journal = {\prc},
         year = 2021,
        month = apr,
       volume = {103},
       number = {4},
          eid = {045810},
        pages = {045810},
          doi = {10.1103/PhysRevC.103.045810},
archivePrefix = {arXiv},
       eprint = {2009.05181},
 primaryClass = {nucl-th},
       adsurl = {https://ui.adsabs.harvard.edu/abs/2021PhRvC.103d5810A}
}

@ARTICLE{1995A&A...297..717Y,
       author = {{Yakovlev}, D.~G. and {Levenfish}, K.~P.},
        title = "{Modified URCA process in neutron star cores.}",
      journal = {\aap},
         year = 1995,
        month = may,
       volume = {297},
        pages = {717},
       adsurl = {https://ui.adsabs.harvard.edu/abs/1995A&A...297..717Y}
}

@article{Lindblom:1999yk,
    author = "Lindblom, Lee and Mendell, Gregory and Owen, Benjamin J.",
    title = "{Second order rotational effects on the r modes of neutron stars}",
    eprint = "gr-qc/9902052",
    archivePrefix = "arXiv",
    doi = "10.1103/PhysRevD.60.064006",
    journal = "Phys. Rev. D",
    volume = "60",
    pages = "064006",
    year = "1999"
}

@article{Lindblom:1998wf,
    author = "Lindblom, Lee and Owen, Benjamin J. and Morsink, Sharon M.",
    title = "{Gravitational radiation instability in hot young neutron stars}",
    eprint = "gr-qc/9803053",
    archivePrefix = "arXiv",
    reportNumber = "GRP-495",
    doi = "10.1103/PhysRevLett.80.4843",
    journal = "Phys. Rev. Lett.",
    volume = "80",
    pages = "4843--4846",
    year = "1998"
}

@article{Haensel:2001mw,
    author = "Haensel, P. and Levenfish, K. P. and Yakovlev, D. G.",
    title = "{Bulk viscosity in superfluid neutron star cores. 2. Modified Urca processes in npe mu matter}",
    eprint = "astro-ph/0103290",
    archivePrefix = "arXiv",
    doi = "10.1051/0004-6361:20010383",
    journal = "Astron. Astrophys.",
    volume = "327",
    pages = "130--137",
    year = "2001"
}

@article{PhysRevD.39.3804,
  title = {Bulk viscosity of hot neutron-star matter and the maximum rotation rates of neutron stars},
  author = {Sawyer, Raymond F.},
  journal = {Phys. Rev. D},
  volume = {39},
  issue = {12},
  pages = {3804--3806},
  numpages = {0},
  year = {1989},
  month = {Jun},
  publisher = {American Physical Society},
  doi = {10.1103/PhysRevD.39.3804},
  url = {https://link.aps.org/doi/10.1103/PhysRevD.39.3804}
}

@ARTICLE{2004ApJ...605..830C,
       author = {{Chang}, Philip and {Bildsten}, Lars},
        title = "{Evolution of Young Neutron Star Envelopes}",
      journal = {\apj},
         year = 2004,
        month = apr,
       volume = {605},
        pages = {830-839},
          doi = {10.1086/382271},
archivePrefix = {arXiv},
       eprint = {astro-ph/0312589},
 primaryClass = {astro-ph},
       adsurl = {https://ui.adsabs.harvard.edu/abs/2004ApJ...605..830C}
}

@article{ParticleDataGroup:2024cfk,
    author = "Navas, S. and others",
    collaboration = "Particle Data Group",
    title = "{Review of particle physics}",
    doi = "10.1103/PhysRevD.110.030001",
    journal = "Phys. Rev. D",
    volume = "110",
    pages = "030001",
    year = "2024"
}

@article{Rezzolla:2017aly,
    author = "Rezzolla, Luciano and Most, Elias R. and Weih, Lukas R.",
    title = "{Using gravitational-wave observations and quasi-universal relations to constrain the maximum mass of neutron stars}",
    eprint = "1711.00314",
    archivePrefix = "arXiv",
    primaryClass = "astro-ph.HE",
    doi = "10.3847/2041-8213/aaa401",
    journal = "Astrophys. J. Lett.",
    volume = "852",
    pages = "L25",
    year = "2018"
}

@article{Huang:2009ue,
    author = "Huang, Xu-Guang and Huang, Mei and Rischke, Dirk H. and Sedrakian, Armen",
    title = "{Anisotropic Hydrodynamics, Bulk Viscosities and R-Modes of Strange Quark Stars with Strong Magnetic Fields}",
    eprint = "0910.3633",
    archivePrefix = "arXiv",
    primaryClass = "astro-ph.HE",
    doi = "10.1103/PhysRevD.81.045015",
    journal = "Phys. Rev. D",
    volume = "81",
    pages = "045015",
    year = "2010"
}

@article{Brown:2002rf,
    author = "Brown, Edward F. and Bildsten, Lars and Chang, Philip",
    title = "{Variability in the thermal emission from accreting neutron star transients}",
    eprint = "astro-ph/0204102",
    archivePrefix = "arXiv",
    doi = "10.1086/341066",
    journal = "Astrophys. J.",
    volume = "574",
    pages = "920",
    year = "2002"
}

@article{vanDalen:2003uy,
    author = "van Dalen, E. N. E. and Dieperink, A. E. L.",
    title = "{Bulk viscosity in neutron stars from hyperons}",
    eprint = "nucl-th/0311103",
    archivePrefix = "arXiv",
    reportNumber = "KVI-1632",
    doi = "10.1103/PhysRevC.69.025802",
    journal = "Phys. Rev. C",
    volume = "69",
    pages = "025802",
    year = "2004"
}

@article{Nayyar:2005th,
    author = "Nayyar, Mohit and Owen, Benjamin J.",
    title = "{R-modes of accreting hyperon stars as persistent sources of gravitational waves}",
    eprint = "astro-ph/0512041",
    archivePrefix = "arXiv",
    reportNumber = "IGPG-05-12-1",
    doi = "10.1103/PhysRevD.73.084001",
    journal = "Phys. Rev. D",
    volume = "73",
    pages = "084001",
    year = "2006"
}

@ARTICLE{2008arXiv0806.3359S,
       author = {{Sa'd}, Basil A.},
        title = "{Quark contribution to r-mode instabilities for several phases of deconfined quark matter}",
      journal = {arXiv e-prints},
         year = 2008,
        month = jun,
          eid = {arXiv:0806.3359},
        pages = {arXiv:0806.3359},
          doi = {10.48550/arXiv.0806.3359},
archivePrefix = {arXiv},
       eprint = {0806.3359},
 primaryClass = {astro-ph},
       adsurl = {https://ui.adsabs.harvard.edu/abs/2008arXiv0806.3359S}
}

@ARTICLE{2001PhRvL..86.1384J,
       author = {{Jones}, P.~B.},
        title = "{Comment on ``Gravitational Radiation Instability in Hot Young Neutron Stars''}",
      journal = {\prl},
         year = 2001,
        month = feb,
       volume = {86},
        pages = {1384-1384},
          doi = {10.1103/PhysRevLett.86.1384},
       adsurl = {https://ui.adsabs.harvard.edu/abs/2001PhRvL..86.1384J}
}

@ARTICLE{2001PhRvD..64h4003J,
       author = {{Jones}, P.~B.},
        title = "{Bulk viscosity of neutron-star matter}",
      journal = {\prd},
         year = 2001,
        month = oct,
       volume = {64},
          eid = {084003},
        pages = {084003},
          doi = {10.1103/PhysRevD.64.084003},
       adsurl = {https://ui.adsabs.harvard.edu/abs/2001PhRvD..64h4003J}
}

@ARTICLE{1969Ap&SS...5..213L,
       author = {{Langer}, William D. and {Cameron}, A.~G.~W.},
        title = "{Effects of Hyperons on the Vibrations of Neutron Stars}",
      journal = {Astrophysics and Space Science},
         year = 1969,
        month = oct,
       volume = {5},
        pages = {213-253},
          doi = {10.1007/BF00650292},
       adsurl = {https://ui.adsabs.harvard.edu/abs/1969Ap&SS...5..213L}
}

@article{Haensel:2001em,
    author = "Haensel, P. and Levenfish, K. P. and Yakovlev, D. G.",
    title = "{Bulk viscosity in superfluid neutron star cores. 3. Effects of sigma- hyperons}",
    eprint = "astro-ph/0110575",
    archivePrefix = "arXiv",
    doi = "10.1051/0004-6361:20011532",
    journal = "Astron. Astrophys.",
    volume = "381",
    pages = "1080--1089",
    year = "2002"
}

@article{Haensel:2000vz,
    author = "Haensel, P. and Levenfish, K. P. and Yakovlev, D. G.",
    title = "{Bulk viscosity in superfluid neutron star cores. I. direct urca processes in npe mu matter}",
    eprint = "astro-ph/0004183",
    archivePrefix = "arXiv",
    url = {https://arxiv.org/abs/astro-ph/0004183},
    journal = "Astron. Astrophys.",
    volume = "357",
    pages = "1157--1169",
    year = "2000"
}

@ARTICLE{1992ApJ...390L..77P,
       author = {{Prakash}, Madappa and {Prakash}, Manju and {Lattimer}, James M. and {Pethick}, C.~J.},
        title = "{Rapid Cooling of Neutron Stars by Hyperons and Delta Isobars}",
      journal = {\apjl},
         year = 1992,
        month = may,
       volume = {390},
        pages = {L77},
          doi = {10.1086/186376},
       adsurl = {https://ui.adsabs.harvard.edu/abs/1992ApJ...390L..77P}
}

@article{Glendenning:1991es,
    author = "Glendenning, N. K. and Moszkowski, S. A.",
    title = "{Reconciliation of neutron star masses and binding of the lambda in hypernuclei}",
    reportNumber = "LBL-30645",
    doi = "10.1103/PhysRevLett.67.2414",
    journal = "Phys. Rev. Lett.",
    volume = "67",
    pages = "2414--2417",
    year = "1991"
}

@article{LIGOScientific:2017vwq,
    author = "Abbott, B. P. and others",
    collaboration = "LIGO Scientific, Virgo",
    title = "{GW170817: Observation of Gravitational Waves from a Binary Neutron Star Inspiral}",
    eprint = "1710.05832",
    archivePrefix = "arXiv",
    primaryClass = "gr-qc",
    reportNumber = "LIGO-P170817",
    doi = "10.1103/PhysRevLett.119.161101",
    journal = "Phys. Rev. Lett.",
    volume = "119",
    pages = "161101",
    year = "2017"
}

@article{Drout:2017ijr,
    author = "Drout, M. R. and others",
    title = "{Light Curves of the Neutron Star Merger GW170817/SSS17a: Implications for R-Process Nucleosynthesis}",
    eprint = "1710.05443",
    archivePrefix = "arXiv",
    primaryClass = "astro-ph.HE",
    doi = "10.1126/science.aaq0049",
    journal = "Science",
    volume = "358",
    pages = "1570--1574",
    year = "2017"
}

@article{Cowperthwaite:2017dyu,
    author = "Cowperthwaite, P. S. and others",
    title = "{The Electromagnetic Counterpart of the Binary Neutron Star Merger LIGO/Virgo GW170817. II. UV, Optical, and Near-infrared Light Curves and Comparison to Kilonova Models}",
    eprint = "1710.05840",
    archivePrefix = "arXiv",
    primaryClass = "astro-ph.HE",
    reportNumber = "FERMILAB-PUB-17-470-A-AE-CD-PPD",
    doi = "10.3847/2041-8213/aa8fc7",
    journal = "Astrophys. J. Lett.",
    volume = "848",
    pages = "L17",
    year = "2017"
}

@article{Annala:2017llu,
    author = "Annala, Eemeli and Gorda, Tyler and Kurkela, Aleksi and Vuorinen, Aleksi",
    title = "{Gravitational-wave constraints on the neutron-star-matter Equation of State}",
    eprint = "1711.02644",
    archivePrefix = "arXiv",
    primaryClass = "astro-ph.HE",
    reportNumber = "CERN-TH-2017-236",
    doi = "10.1103/PhysRevLett.120.172703",
    journal = "Phys. Rev. Lett.",
    volume = "120",
    pages = "172703",
    year = "2018"
}

@article{Bauswein:2017vtn,
    author = "Bauswein, Andreas and Just, Oliver and Janka, Hans-Thomas and Stergioulas, Nikolaos",
    title = "{Neutron-star radius constraints from GW170817 and future detections}",
    eprint = "1710.06843",
    archivePrefix = "arXiv",
    primaryClass = "astro-ph.HE",
    doi = "10.3847/2041-8213/aa9994",
    journal = "Astrophys. J. Lett.",
    volume = "850",
    pages = "L34",
    year = "2017"
}

@article{Shibata:2019ctb,
    author = "Shibata, Masaru and Zhou, Enping and Kiuchi, Kenta and Fujibayashi, Sho",
    title = "{Constraint on the maximum mass of neutron stars using GW170817 event}",
    eprint = "1905.03656",
    archivePrefix = "arXiv",
    primaryClass = "astro-ph.HE",
    doi = "10.1103/PhysRevD.100.023015",
    journal = "Phys. Rev. D",
    volume = "100",
    pages = "023015",
    year = "2019"
}

@article{Margalit_2017,
   title={Constraining the Maximum Mass of Neutron Stars from Multi-messenger Observations of GW170817},
   volume={850},
   ISSN={2041-8213},
   url={http://dx.doi.org/10.3847/2041-8213/aa991c},
   DOI={10.3847/2041-8213/aa991c},
   journal={The Astrophysical Journal Letters},
   publisher={American Astronomical Society},
   author={Margalit, Ben and Metzger, Brian D.},
   year={2017},
   month=nov, pages={L19} }

@article{Radice_2018,
   title={GW170817: Joint Constraint on the Neutron Star Equation of State from Multimessenger Observations},
   volume={852},
   ISSN={2041-8213},
   url={http://dx.doi.org/10.3847/2041-8213/aaa402},
   DOI={10.3847/2041-8213/aaa402},
   journal={The Astrophysical Journal Letters},
   publisher={American Astronomical Society},
   author={Radice, David and Perego, Albino and Zappa, Francesco and Bernuzzi, Sebastiano},
   year={2018},
   month=jan, pages={L29} }

@article{Ruiz:2017due,
    author = "Ruiz, Milton and Shapiro, Stuart L. and Tsokaros, Antonios",
    title = "{GW170817, General Relativistic Magnetohydrodynamic Simulations, and the Neutron Star Maximum Mass}",
    eprint = "1711.00473",
    archivePrefix = "arXiv",
    primaryClass = "astro-ph.HE",
    doi = "10.1103/PhysRevD.97.021501",
    journal = "Phys. Rev. D",
    volume = "97",
    pages = "021501",
    year = "2018"
}

@article{Most:2018hfd,
    author = {Most, Elias R. and Weih, Lukas R. and Rezzolla, Luciano and Schaffner-Bielich, J{\"u}rgen},
    title = "{New constraints on radii and tidal deformabilities of neutron stars from GW170817}",
    eprint = "1803.00549",
    archivePrefix = "arXiv",
    primaryClass = "gr-qc",
    doi = "10.1103/PhysRevLett.120.261103",
    journal = "Phys. Rev. Lett.",
    volume = "120",
    pages = "261103",
    year = "2018"
}

@article{Raithel:2019uzi,
    author = "Raithel, Carolyn A.",
    title = "{Constraints on the Neutron Star Equation of State from GW170817}",
    eprint = "1904.10002",
    archivePrefix = "arXiv",
    primaryClass = "astro-ph.HE",
    doi = "10.1140/epja/i2019-12759-5",
    journal = "Eur. Phys. J. A",
    volume = "55",
    pages = "80",
    year = "2019"
}

@article{Shibata_2019,
   title={Constraint on the maximum mass of neutron stars using GW170817 event},
   volume={100},
   ISSN={2470-0029},
   url={http://dx.doi.org/10.1103/PhysRevD.100.023015},
   DOI={10.1103/physrevd.100.023015},
   journal={Physical Review D},
   publisher={American Physical Society (APS)},
   author={Shibata, Masaru and Zhou, Enping and Kiuchi, Kenta and Fujibayashi, Sho},
   year={2019},
   month=jul }

@article{Dietrich:2020efo,
    author = "Dietrich, Tim and Coughlin, Michael W. and Pang, Peter T. H. and Bulla, Mattia and Heinzel, Jack and Issa, Lina and Tews, Ingo and Antier, Sarah",
    title = "{Multimessenger constraints on the neutron-star equation of state and the Hubble constant}",
    eprint = "2002.11355",
    archivePrefix = "arXiv",
    primaryClass = "astro-ph.HE",
    reportNumber = "LA-UR-20-21470",
    doi = "10.1126/science.abb4317",
    journal = "Science",
    volume = "370",
    pages = "1450--1453",
    year = "2020"
}

@article{Nathanail:2021tay,
    author = "Nathanail, Antonios and Most, Elias R. and Rezzolla, Luciano",
    title = "{GW170817 and GW190814: tension on the maximum mass}",
    eprint = "2101.01735",
    archivePrefix = "arXiv",
    primaryClass = "astro-ph.HE",
    doi = "10.3847/2041-8213/abdfc6",
    journal = "Astrophys. J. Lett.",
    volume = "908",
    pages = "L28",
    year = "2021"
}

@article{Stergioulas_2011,
   title={Gravitational waves and non-axisymmetric oscillation modes in mergers of compact object binaries: Gravitational waves and oscillation modes},
   volume={418},
   ISSN={0035-8711},
   url={http://dx.doi.org/10.1111/j.1365-2966.2011.19493.x},
   DOI={10.1111/j.1365-2966.2011.19493.x},
   journal={Monthly Notices of the Royal Astronomical Society},
   publisher={Oxford University Press (OUP)},
   author={Stergioulas, Nikolaos and Bauswein, Andreas and Zagkouris, Kimon and Janka, Hans-Thomas},
   year={2011},
   month=sep, pages={427–436} }

@article{Bauswein:2011tp,
    author = "Bauswein, A. and Janka, H. -Th.",
    title = "{Measuring neutron-star properties via gravitational waves from binary mergers}",
    eprint = "1106.1616",
    archivePrefix = "arXiv",
    primaryClass = "astro-ph.SR",
    doi = "10.1103/PhysRevLett.108.011101",
    journal = "Phys. Rev. Lett.",
    volume = "108",
    pages = "011101",
    year = "2012"
}

@article{Takami:2014zpa,
    author = "Takami, Kentaro and Rezzolla, Luciano and Baiotti, Luca",
    title = "{Constraining the Equation of State of Neutron Stars from Binary Mergers}",
    eprint = "1403.5672",
    archivePrefix = "arXiv",
    primaryClass = "gr-qc",
    doi = "10.1103/PhysRevLett.113.091104",
    journal = "Phys. Rev. Lett.",
    volume = "113",
    pages = "091104",
    year = "2014"
}

@article{Bernuzzi:2015rla,
    author = "Bernuzzi, Sebastiano and Dietrich, Tim and Nagar, Alessandro",
    title = "{Modeling the complete gravitational wave spectrum of neutron star mergers}",
    eprint = "1504.01764",
    archivePrefix = "arXiv",
    primaryClass = "gr-qc",
    doi = "10.1103/PhysRevLett.115.091101",
    journal = "Phys. Rev. Lett.",
    volume = "115",
    pages = "091101",
    year = "2015"
}

@article{PhysRevD.93.124051,
  title = {Gravitational-wave signal from binary neutron stars: A systematic analysis of the spectral properties},
  author = {Rezzolla, Luciano and Takami, Kentaro},
  journal = {Phys. Rev. D},
  volume = {93},
  issue = {12},
  pages = {124051},
  numpages = {20},
  year = {2016},
  month = {Jun},
  publisher = {American Physical Society},
  doi = {10.1103/PhysRevD.93.124051},
  url = {https://link.aps.org/doi/10.1103/PhysRevD.93.124051}
}

@article{PhysRevLett.128.161102,
  title = {Constraints on the Maximum Densities of Neutron Stars from Postmerger Gravitational Waves with Third-Generation Observations},
  author = {Breschi, Matteo and Bernuzzi, Sebastiano and Godzieba, Daniel and Perego, Albino and Radice, David},
  journal = {Phys. Rev. Lett.},
  volume = {128},
  issue = {16},
  pages = {161102},
  numpages = {7},
  year = {2022},
  month = {Apr},
  publisher = {American Physical Society},
  doi = {10.1103/PhysRevLett.128.161102},
  url = {https://link.aps.org/doi/10.1103/PhysRevLett.128.161102}
}

@article{Most:2018eaw,
    author = {Most, Elias R. and Papenfort, L. Jens and Dexheimer, Veronica and Hanauske, Matthias and Schramm, Stefan and St{\"o}cker, Horst and Rezzolla, Luciano},
    title = "{Signatures of quark-hadron phase transitions in general-relativistic neutron-star mergers}",
    eprint = "1807.03684",
    archivePrefix = "arXiv",
    primaryClass = "astro-ph.HE",
    doi = "10.1103/PhysRevLett.122.061101",
    journal = "Phys. Rev. Lett.",
    volume = "122",
    pages = "061101",
    year = "2019"
}

@article{Most_2020,
   title={On the deconfinement phase transition in neutron-star mergers},
   volume={56},
   ISSN={1434-601X},
   url={http://dx.doi.org/10.1140/epja/s10050-020-00073-4},
   DOI={10.1140/epja/s10050-020-00073-4},
   number={2},
   journal={The European Physical Journal A},
   publisher={Springer Science and Business Media LLC},
   author={Most, Elias R. and Jens Papenfort, L. and Dexheimer, Veronica and Hanauske, Matthias and Stoecker, Horst and Rezzolla, Luciano},
   year={2020},
   month=feb }

@article{Weih:2019xvw,
    author = "Weih, Lukas R. and Hanauske, Matthias and Rezzolla, Luciano",
    title = "{Postmerger Gravitational-Wave Signatures of Phase Transitions in Binary Mergers}",
    eprint = "1912.09340",
    archivePrefix = "arXiv",
    primaryClass = "gr-qc",
    doi = "10.1103/PhysRevLett.124.171103",
    journal = "Phys. Rev. Lett.",
    volume = "124",
    pages = "171103",
    year = "2020"
}

@article{Tootle:2022pvd,
    author = {Tootle, Samuel and Ecker, Christian and Topolski, Konrad and Demircik, Tuna and J{\"a}rvinen, Matti and Rezzolla, Luciano},
    title = "{Quark formation and phenomenology in binary neutron-star mergers using V-QCD}",
    eprint = "2205.05691",
    archivePrefix = "arXiv",
    primaryClass = "astro-ph.HE",
    reportNumber = "APCTP Pre2022 - 007",
    doi = "10.21468/SciPostPhys.13.5.109",
    journal = "SciPost Phys.",
    volume = "13",
    pages = "109",
    year = "2022"
}

@article{Bauswein:2018bma,
    author = "Bauswein, Andreas and Bastian, Niels-Uwe F. and Blaschke, David B. and Chatziioannou, Katerina and Clark, James A. and Fischer, Tobias and Oertel, Micaela",
    title = "{Identifying a first-order phase transition in neutron star mergers through gravitational waves}",
    eprint = "1809.01116",
    archivePrefix = "arXiv",
    primaryClass = "astro-ph.HE",
    doi = "10.1103/PhysRevLett.122.061102",
    journal = "Phys. Rev. Lett.",
    volume = "122",
    pages = "061102",
    year = "2019"
}

@article{Blacker:2020nlq,
    author = "Blacker, Sebastian and Bastian, Niels-Uwe F. and Bauswein, Andreas and Blaschke, David B. and Fischer, Tobias and Oertel, Micaela and Soultanis, Theodoros and Typel, Stefan",
    title = "{Constraining the onset density of the hadron-quark phase transition with gravitational-wave observations}",
    eprint = "2006.03789",
    archivePrefix = "arXiv",
    primaryClass = "astro-ph.HE",
    doi = "10.1103/PhysRevD.102.123023",
    journal = "Phys. Rev. D",
    volume = "102",
    pages = "123023",
    year = "2020"
}

@article{Liebling:2020dhf,
    author = "Liebling, Steven L. and Palenzuela, Carlos and Lehner, Luis",
    title = "{Effects of High Density Phase Transitions on Neutron Star Dynamics}",
    eprint = "2010.12567",
    archivePrefix = "arXiv",
    primaryClass = "gr-qc",
    doi = "10.1088/1361-6382/abf898",
    journal = "Class. Quant. Grav.",
    volume = "38",
    pages = "115007",
    year = "2021"
}

@article{Prakash:2021wpz,
    author = "Prakash, Aviral and Radice, David and Logoteta, Domenico and Perego, Albino and Nedora, Vsevolod and Bombaci, Ignazio and Kashyap, Rahul and Bernuzzi, Sebastiano and Endrizzi, Andrea",
    title = "{Signatures of deconfined quark phases in binary neutron star mergers}",
    eprint = "2106.07885",
    archivePrefix = "arXiv",
    primaryClass = "astro-ph.HE",
    doi = "10.1103/PhysRevD.104.083029",
    journal = "Phys. Rev. D",
    volume = "104",
    pages = "083029",
    year = "2021"
}

@article{Fujimoto:2022xhv,
    author = "Fujimoto, Yuki and Fukushima, Kenji and Hotokezaka, Kenta and Kyutoku, Koutarou",
    title = "{Gravitational Wave Signal for Quark Matter with Realistic Phase Transition}",
    eprint = "2205.03882",
    archivePrefix = "arXiv",
    primaryClass = "astro-ph.HE",
    reportNumber = "INT-PUB-22-015",
    doi = "10.1103/PhysRevLett.130.091404",
    journal = "Phys. Rev. Lett.",
    volume = "130",
    pages = "091404",
    year = "2023"
}

@article{Ujevic:2022nkr,
    author = "Ujevic, Maximiliano and Gieg, Henrique and Schianchi, Federico and Chaurasia, Swami Vivekanandji and Tews, Ingo and Dietrich, Tim",
    title = "{Reverse phase transitions in binary neutron-star systems with exotic-matter cores}",
    eprint = "2211.04662",
    archivePrefix = "arXiv",
    primaryClass = "gr-qc",
    reportNumber = "Report-no: LA-UR-22-31740",
    doi = "10.1103/PhysRevD.107.024025",
    journal = "Phys. Rev. D",
    volume = "107",
    pages = "024025",
    year = "2023"
}

@article{Godzieba:2020tjn,
    author = "Godzieba, Daniel A. and Radice, David and Bernuzzi, Sebastiano",
    title = "{On the maximum mass of neutron stars and GW190814}",
    eprint = "2007.10999",
    archivePrefix = "arXiv",
    primaryClass = "astro-ph.HE",
    doi = "10.3847/1538-4357/abd4dd",
    journal = "Astrophys. J.",
    volume = "908",
    pages = "122",
    year = "2021"
}

@article{Alford:2017rxf,
    author = "Alford, Mark G. and Bovard, Luke and Hanauske, Matthias and Rezzolla, Luciano and Schwenzer, Kai",
    title = "{Viscous Dissipation and Heat Conduction in Binary Neutron-Star Mergers}",
    eprint = "1707.09475",
    archivePrefix = "arXiv",
    primaryClass = "gr-qc",
    doi = "10.1103/PhysRevLett.120.041101",
    journal = "Phys. Rev. Lett.",
    volume = "120",
    pages = "041101",
    year = "2018"
}

@article{alford2021bulk,
  title={Bulk viscosity from Urca processes: npe $\mu$ matter in the neutrino-trapped regime},
  author={Alford, Mark and Harutyunyan, Arus and Sedrakian, Armen},
  journal={Physical Review D},
  volume={104},
  pages={103027},
  year={2021},
  publisher={APS}
}

@article{PhysRevD.100.103021,
  title = {Bulk viscosity of baryonic matter with trapped neutrinos},
  author = {Alford, Mark and Harutyunyan, Arus and Sedrakian, Armen},
  journal = {Phys. Rev. D},
  volume = {100},
  issue = {10},
  pages = {103021},
  numpages = {26},
  year = {2019},
  month = {Nov},
  publisher = {American Physical Society},
  doi = {10.1103/PhysRevD.100.103021},
  url = {https://link.aps.org/doi/10.1103/PhysRevD.100.103021}
}

@article{Alford:2023gxq,
    author = "Alford, Mark G. and Haber, Alexander and Zhang, Ziyuan",
    title = "{Isospin equilibration in neutron star mergers}",
    eprint = "2306.06180",
    archivePrefix = "arXiv",
    primaryClass = "nucl-th",
    doi = "10.1103/PhysRevC.109.055803",
    journal = "Phys. Rev. C",
    volume = "109",
    pages = "055803",
    year = "2024"
}

@article{PhysRevC.109.015805,
  title = {Far-from-equilibrium bulk-viscous transport coefficients in neutron star mergers},
  author = {Yang, Yumu and Hippert, Mauricio and Speranza, Enrico and Noronha, Jorge},
  journal = {Phys. Rev. C},
  volume = {109},
  issue = {1},
  pages = {015805},
  numpages = {19},
  year = {2024},
  month = {Jan},
  publisher = {American Physical Society},
  doi = {10.1103/PhysRevC.109.015805},
  url = {https://link.aps.org/doi/10.1103/PhysRevC.109.015805}
}

@article{Alford:2021lpp,
    author = "Alford, Mark and Harutyunyan, Arus and Sedrakian, Armen",
    title = "{Bulk viscosity from Urca processes: npe{\ensuremath{\mu}} matter in the neutrino-trapped regime}",
    eprint = "2108.07523",
    archivePrefix = "arXiv",
    primaryClass = "astro-ph.HE",
    doi = "10.1103/PhysRevD.104.103027",
    journal = "Phys. Rev. D",
    volume = "104",
    pages = "103027",
    year = "2021"
}

@article{PhysRevC.100.035803,
  title = {Damping of density oscillations in neutrino-transparent nuclear matter},
  author = {Alford, Mark G. and Harris, Steven P.},
  journal = {Phys. Rev. C},
  volume = {100},
  issue = {3},
  pages = {035803},
  numpages = {13},
  year = {2019},
  month = {Sep},
  publisher = {American Physical Society},
  doi = {10.1103/PhysRevC.100.035803},
  url = {https://link.aps.org/doi/10.1103/PhysRevC.100.035803}
}

@article{Alford:2024tyj,
    author = "Alford, Mark and Harutyunyan, Arus and Sedrakian, Armen and Tsiopelas, Stefanos",
    title = "{Bulk viscosity of two-color superconducting quark matter in neutron star mergers}",
    eprint = "2407.12493",
    archivePrefix = "arXiv",
    primaryClass = "nucl-th",
    doi = "10.1103/PhysRevD.110.L061303",
    journal = "Phys. Rev. D",
    volume = "110",
    pages = "L061303",
    year = "2024"
}

@article{CruzRojas:2024etx,
    author = {Cruz Rojas, Jes{\'u}s and Gorda, Tyler and Hoyos, Carlos and Jokela, Niko and J{\"a}rvinen, Matti and Kurkela, Aleksi and Paatelainen, Risto and S{\"a}ppi, Saga and Vuorinen, Aleksi},
    title = "{Estimate for the Bulk Viscosity of Strongly Coupled Quark Matter Using Perturbative QCD and Holography}",
    eprint = "2402.00621",
    archivePrefix = "arXiv",
    primaryClass = "hep-ph",
    reportNumber = "APCTP Pre2024 - 003, HIP-2024-1/TH, TUM-EFT 187/24",
    doi = "10.1103/PhysRevLett.133.071901",
    journal = "Phys. Rev. Lett.",
    volume = "133",
    pages = "071901",
    year = "2024"
}

@article{Hern_ndez_2024,
   title={Damping of density oscillations from bulk viscosity in quark matter},
   volume={109},
   ISSN={2470-0029},
   url={http://dx.doi.org/10.1103/PhysRevD.109.123022},
   DOI={10.1103/physrevd.109.123022},
   journal={Physical Review D},
   publisher={American Physical Society (APS)},
   author={Hernández, José Luis and Manuel, Cristina and Tolos, Laura},
   year={2024},
   month=jun }

@article{PhysRevD.107.103032,
  title = {Simulating bulk viscosity in neutron stars. II. Evolution in spherical symmetry},
  author = {Camelio, Giovanni and Gavassino, Lorenzo and Antonelli, Marco and Bernuzzi, Sebastiano and Haskell, Brynmor},
  journal = {Phys. Rev. D},
  volume = {107},
  issue = {10},
  pages = {103032},
  numpages = {23},
  year = {2023},
  month = {May},
  publisher = {American Physical Society},
  doi = {10.1103/PhysRevD.107.103032},
  url = {https://link.aps.org/doi/10.1103/PhysRevD.107.103032}
}

@article{Most_2024,
   title={Emergence of Microphysical Bulk Viscosity in Binary Neutron Star Postmerger Dynamics},
   volume={967},
   ISSN={2041-8213},
   url={http://dx.doi.org/10.3847/2041-8213/ad454f},
   DOI={10.3847/2041-8213/ad454f},
   number={1},
   journal={The Astrophysical Journal Letters},
   publisher={American Astronomical Society},
   author={Most, Elias R. and Haber, Alexander and Harris, Steven P. and Zhang, Ziyuan and Alford, Mark G. and Noronha, Jorge},
   year={2024},
   month=may, pages={L14} }

@article{Chabanov_2025,
   title={Impact of Bulk Viscosity on the Postmerger Gravitational-Wave Signal from Merging Neutron Stars},
   volume={134},
   ISSN={1079-7114},
   url={http://dx.doi.org/10.1103/PhysRevLett.134.071402},
   DOI={10.1103/physrevlett.134.071402},
   journal={Physical Review Letters},
   publisher={American Physical Society (APS)},
   author={Chabanov, Michail and Rezzolla, Luciano},
   year={2025},
   month=feb }

@article{Espino:2023dei,
    author = "Espino, Pedro Luis and Hammond, Peter and Radice, David and Bernuzzi, Sebastiano and Gamba, Rossella and Zappa, Francesco and Longo Micchi, Luis Felipe and Perego, Albino",
    title = "{Neutrino Trapping and Out-of-Equilibrium Effects in Binary Neutron-Star Merger Remnants}",
    eprint = "2311.00031",
    archivePrefix = "arXiv",
    primaryClass = "astro-ph.HE",
    doi = "10.1103/PhysRevLett.132.211001",
    journal = "Phys. Rev. Lett.",
    volume = "132",
    pages = "211001",
    year = "2024"
}

@article{Chabanov:2023abq,
    author = "Chabanov, Michail and Rezzolla, Luciano",
    title = "{Numerical modeling of bulk viscosity in neutron stars}",
    eprint = "2311.13027",
    archivePrefix = "arXiv",
    primaryClass = "gr-qc",
    doi = "10.1103/PhysRevD.111.044074",
    journal = "Phys. Rev. D",
    volume = "111",
    pages = "044074",
    year = "2025"
}

@article{Radice_2022,
   title={A new moment-based general-relativistic neutrino-radiation transport code: Methods and first applications to neutron star mergers},
   volume={512},
   ISSN={1365-2966},
   url={http://dx.doi.org/10.1093/mnras/stac589},
   DOI={10.1093/mnras/stac589},
   journal={Monthly Notices of the Royal Astronomical Society},
   publisher={Oxford University Press (OUP)},
   author={Radice, David and Bernuzzi, Sebastiano and Perego, Albino and Haas, Roland},
   year={2022},
   month=mar, pages={1499–1521} }

@article{Zappa_2023,
   title={Binary neutron star merger simulations with neutrino transport and turbulent viscosity: impact of different schemes and grid resolution},
   volume={520},
   ISSN={1365-2966},
   url={http://dx.doi.org/10.1093/mnras/stad107},
   DOI={10.1093/mnras/stad107},
   journal={Monthly Notices of the Royal Astronomical Society},
   publisher={Oxford University Press (OUP)},
   author={Zappa, Francesco and Bernuzzi, Sebastiano and Radice, David and Perego, Albino},
   year={2023},
   month=jan, pages={1481–1503} }

@article{Most_2021,
   title={Projecting the likely importance of weak-interaction-driven bulk viscosity in neutron star mergers},
   volume={509},
   ISSN={1365-2966},
   url={http://dx.doi.org/10.1093/mnras/stab2793},
   DOI={10.1093/mnras/stab2793},
   journal={Monthly Notices of the Royal Astronomical Society},
   publisher={Oxford University Press (OUP)},
   author={Most, Elias R and Harris, Steven P and Plumberg, Christopher and Alford, Mark G and Noronha, Jorge and Noronha-Hostler, Jacquelyn and Pretorius, Frans and Witek, Helvi and Yunes, Nicolás},
   year={2021},
   month=oct, pages={1096–1108} }

@article{Kraav_2024,
   title={Instability windows of relativistic 
<mml:math xmlns:mml="http://www.w3.org/1998/Math/MathML" display="inline"><mml:mi>r</mml:mi></mml:math>
-modes},
   volume={109},
   ISSN={2470-0029},
   url={http://dx.doi.org/10.1103/PhysRevD.109.043012},
   DOI={10.1103/physrevd.109.043012},
   journal={Physical Review D},
   publisher={American Physical Society (APS)},
   author={Kraav, Kirill Y. and Gusakov, Mikhail E. and Kantor, Elena M.},
   year={2024},
   month=feb }

@article{ANDERSSON_2001,
   title={THE R-MODE INSTABILITY IN ROTATING NEUTRON STARS},
   volume={10},
   ISSN={1793-6594},
   url={http://dx.doi.org/10.1142/S0218271801001062},
   DOI={10.1142/s0218271801001062},
   journal={International Journal of Modern Physics D},
   publisher={World Scientific Pub Co Pte Lt},
   author={ANDERSSON, NILS and KOKKOTAS, KOSTAS D.},
   year={2001},
   month=aug, pages={381–441} }

@article{Li:2023gtg,
    author = "Li, Xiyuan and Abbassi, Shahram and Upadhyaya, Varenya and Zhang, Xiyang and Valluri, S. R.",
    title = "{The role of r-modes in pulsar spin-down, pulsar timing, and gravitational waves}",
    eprint = "2307.11270",
    archivePrefix = "arXiv",
    primaryClass = "astro-ph.HE",
    doi = "10.1016/j.jheap.2025.100446",
    journal = "JHEAp",
    volume = "49",
    pages = "100446",
    year = "2026"
}

@article{Sinha:2008wb,
    author = "Sinha, Monika and Bandyopadhyay, Debades",
    title = "{Hyperon bulk viscosity in strong magnetic fields}",
    eprint = "0809.3337",
    archivePrefix = "arXiv",
    primaryClass = "astro-ph",
    doi = "10.1103/PhysRevD.79.123001",
    journal = "Phys. Rev. D",
    volume = "79",
    pages = "123001",
    year = "2009"
}

@article{Dexheimer_2021,
   title={GW190814 as a massive rapidly rotating neutron star with exotic degrees of freedom},
   volume={103},
   ISSN={2469-9993},
   url={http://dx.doi.org/10.1103/PhysRevC.103.025808},
   DOI={10.1103/physrevc.103.025808},
   journal={Physical Review C},
   publisher={American Physical Society (APS)},
   author={Dexheimer, V. and Gomes, R. O. and Klähn, T. and Han, S. and Salinas, M.},
   year={2021},
   month=feb }

@article{Biswas:2020xna,
    author = "Biswas, Bhaskar and Nandi, Rana and Char, Prasanta and Bose, Sukanta and Stergioulas, Nikolaos",
    title = "{GW190814: on the properties of the secondary component of the binary}",
    eprint = "2010.02090",
    archivePrefix = "arXiv",
    primaryClass = "astro-ph.HE",
    reportNumber = "LIGO preprint number LIGO-P2000368, Virgo preprint number
  VIR-0807B-20",
    doi = "10.1093/mnras/stab1383",
    journal = "Mon. Not. Roy. Astron. Soc.",
    volume = "505",
    pages = "1600--1606",
    year = "2021"
}

@article{Fattoyev:2020cws,
    author = "Fattoyev, F. J. and Horowitz, C. J. and Piekarewicz, J. and Reed, Brendan",
    title = "{GW190814: Impact of a 2.6 solar mass neutron star on the nucleonic equations of state}",
    eprint = "2007.03799",
    archivePrefix = "arXiv",
    primaryClass = "nucl-th",
    doi = "10.1103/PhysRevC.102.065805",
    journal = "Phys. Rev. C",
    volume = "102",
    pages = "065805",
    year = "2020"
}

@article{Zhou_2021,
   title={R-mode Stability of GW190814’s Secondary Component as a Supermassive and Superfast Pulsar},
   volume={910},
   ISSN={1538-4357},
   url={http://dx.doi.org/10.3847/1538-4357/abe538},
   DOI={10.3847/1538-4357/abe538},
   journal={The Astrophysical Journal},
   publisher={American Astronomical Society},
   author={Zhou, Xia and Li, Ang and Li, Bao-An},
   year={2021},
   month=mar, pages={62} }

@misc{papazoglou2017rmodeconstraintsneutronstar,
      title={R-mode constraints from neutron star equation of state}, 
      author={M. C. Papazoglou and C. C. Moustakidis},
      year={2017},
      eprint={1506.04572},
      archivePrefix={arXiv},
      primaryClass={astro-ph.SR},
      url={https://arxiv.org/abs/1506.04572}, 
}

@book{Glendenning:1997wn,
    author = "Glendenning, N. K.",
    title = "{Compact stars: Nuclear physics, particle physics, and general relativity}",
    doi = "10.1007/978-1-4684-0491-3",
    publisher = "Springer", 
    series = "Astronomy and Astrophysics Library",   
    year = "1997"
}

@article{Ai:2023ykc,
    author = "Ai, Shunke and Gao, He and Yuan, Yong and Zhang, Bing and Lan, Lin",
    title = "{What constraints can one pose on the maximum mass of neutron stars from multimessenger observations?}",
    eprint = "2310.07133",
    archivePrefix = "arXiv",
    primaryClass = "astro-ph.HE",
    doi = "10.1093/mnras/stad3177",
    journal = "Mon. Not. Roy. Astron. Soc.",
    volume = "526",
    pages = "6260--6273",
    year = "2023"
}

@article{Margalit:2022rde,
    author = "Margalit, Ben and Jermyn, Adam S. and Metzger, Brian D. and Roberts, Luke F. and Quataert, Eliot",
    title = "{Angular-momentum Transport in Proto-neutron Stars and the Fate of Neutron Star Merger Remnants}",
    eprint = "2206.10645",
    archivePrefix = "arXiv",
    primaryClass = "astro-ph.HE",
    doi = "10.3847/1538-4357/ac8b01",
    journal = "Astrophys. J.",
    volume = "939",
    pages = "51",
    year = "2022"
}

@ARTICLE{1939PhRv...55..364T,
       author = {{Tolman}, Richard C.},
        title = "{Static Solutions of Einstein's Field Equations for Spheres of Fluid}",
      journal = {Physical Review},
         year = 1939,
        month = feb,
       volume = {55},
        pages = {364-373},
          doi = {10.1103/PhysRev.55.364},
       adsurl = {https://ui.adsabs.harvard.edu/abs/1939PhRv...55..364T}
}

@ARTICLE{1939PhRv...55..374O,
       author = {{Oppenheimer}, J.~R. and {Volkoff}, G.~M.},
        title = "{On Massive Neutron Cores}",
      journal = {Physical Review},
         year = 1939,
        month = feb,
       volume = {55},
        pages = {374-381},
          doi = {10.1103/PhysRev.55.374},
       adsurl = {https://ui.adsabs.harvard.edu/abs/1939PhRv...55..374O}
}

@ARTICLE{1997A&A...328..274B,
       author = {{Baldo}, M. and {Bombaci}, I. and {Burgio}, G.~F.},
        title = "{Microscopic nuclear equation of state with three-body forces and neutron star structure}",
      journal = {\aap},
         year = 1997,
        month = dec,
       volume = {328},
        pages = {274-282},
          doi = {10.48550/arXiv.astro-ph/9707277},
archivePrefix = {arXiv},
       eprint = {astro-ph/9707277},
 primaryClass = {astro-ph},
       adsurl = {https://ui.adsabs.harvard.edu/abs/1997A&A...328..274B}
}

\end{document}